\documentclass[a4paper,twocolumn,showpacs,preprintnumbers,amsmath,amssymb]{revtex4} 
\usepackage{hyperref}
\usepackage{amsfonts}
\usepackage{epsf}
\usepackage{t1enc}

\usepackage{graphicx}
\usepackage{dcolumn}
\usepackage{bm}

\DeclareFontFamily{OT1}{rsfs}{} \DeclareFontShape{OT1}{rsfs}{m}{n}{
<-7> rsfs5 <7-10> rsfs7 <10-> rsfs10}{}
\DeclareMathAlphabet{\mycal}{OT1}{rsfs}{m}{n}

\def\scri{{\mycal I}}
\def\scrip{\scri^{+}}%

\begin{document}

\title{ Numerical investigation of highly excited magnetic monopoles
  in $SU(2)$ Yang-Mills--Higgs theory}

\author{Gyula Fodor}
\email{gfodor@rmki.kfki.hu}
\author{Istv\'{a}n R\'{a}cz}
\email{iracz@sunserv.kfki.hu}

\affiliation{%
MTA RMKI, H-1525 Budapest 114, P.O.Box 49, Hungary}%

\date{\today}

\begin{abstract}
Preliminary results concerning the time evolution of strongly exited
$SU(2)$ Bogomolny-Prasad-Sommerfield  (BPS) magnetic monopoles have
been published in \cite{fr2}. The behavior of these dynamical magnetic
monopoles was investigated by means of numerical simulations in the
four dimensional Minkowski spacetime. The developed code incorporates
both the techniques of conformal compactification and that of the
hyperboloidal initial value problem. Our primary aim here is to
provide a detailed account on the methods and results of the
investigations reported in \cite{fr2}. In addition, some important new
results, which go much beyond the scope of these early studies, are
also presented. In particular, to be able to distinguish linear and
non-linear effects, evolutions of monopoles deformed by various
excitations, including both very small and extra large energy
excitation, are investigated.  In addition, a detailed account is
provided on the spacetime dependence of the basic variables, as well
as, that of the other physically significant dynamical qualities such
as the energy and energy current densities, the radial and angular
pressures and the magnetic charge density. A careful comprehensive
study of the associated energy transfers and energy balances is also
included.
\end{abstract}

\pacs{03.50.Kk, 14.80.Hv}
\maketitle



\section{Introduction}

Soliton and quasi-soliton type configurations, which are in general
spatially localized non-singular finite energy solutions of non-linear
field  equations, play significant role in various particle physics
considerations (see e.g. \cite{rub,Sutcliffe-Manton} for recent
reviews). In general the associated field equations are highly
non-linear, therefore, most of the investigations have been restricted
to the study of time independent configurations. Thereby,  there is an
obvious increase of interest to study the dynamical properties
of soliton type configurations.  It is a matter of fact that the
available analytic techniques either have not been developed enough to
provide or simply cannot guarantee completely satisfactory answers to
all questions in case of non-linear systems. Therefore, reliable
numerical approaches  are needed which are  able to describe the time
evolution of these type of systems.  Motivated by this sort of
necessities we have developed a  numerical method which have been used
in studying the time evolution of various spherically symmetric
non-linear dynamical systems  \cite{fr1,fr2,fr3,ffpr} (see also
\cite{csp,csp2,csfr,csfrr}). 
The numerical approach we applied is based on
the ``method of lines'' in a fourth order setting \cite{gko} which was
found to be the most efficient among the numerical methods which are
applied in various considerations \cite{hkn}. In addition, to get a
faithful representation of all the radiation processes, the techniques
of conformal compactification (introduced first by Penrose
\cite{Penrose}) to the underlying Minkowski spacetime,
along with the hyperboloidal initial value problem were
incorporated. The applied conformal  gauge, which is a modification of
the static hyperboloidal gauge, makes it possible to investigate by
numerical means the asymptotic properties of radiation processes. One
of the associated advantages is that by making use of this method the
radiation processes, which last infinitely long in physical time, can
be analyzed in finite computational time intervals.

Among the great variety of physically interesting soliton type
configurations distinguished attention has been paid to the study of
magnetic monopole type configurations such as the 't\,Hooft-Polyakov
magnetic monopole solutions of coupled Yang-Mills--Higgs (YMH) systems
\cite{P}. However, even in the case of magnetic monopoles there was a
lack of knowledge concerning the dynamical properties of these
systems. Therefore, our numerical method was first applied to the
study of the dynamical properties of the simplest possible magnetic
monopole system, which could be chosen so that there was a radiative
component among the basic field variables. This paper  is, in fact, to
report about the results of investigations concerning the time
evolution of a strongly excited spherically symmetric $SU(2)$ BPS
magnetic monopole \cite{bo}.

To have a clear enough setting, at least from an ``energetic'' point
of view, we investigated the time evolution of an initially static
SU(2) BPS magnetic monopole which was excited by the help of a high
energy pulse. The self-interaction of the Higgs field was turned off
which yielded a Yang-Mills--Higgs system so that the Yang-Mills field
is massive while the Higgs field is massless.

The dynamics started by injecting a pulse of excitation via the time
derivative of the Yang-Mills variable. Since the Yang-Mills and Higgs
variables are coupled (in a non-linear manner) energy was transformed
to the massless Higgs field which immediately forwarded 
{ {\it about half}} of the energy of the pulse towards future null infinity,
$\scrip$. { It is already a surprise that this reaction of the monopole
is apparently independent of the energy content of the
exciting pulse, at least in case of the subclass considered in this
paper (see subsection \ref{enentr}). Nevertheless, naively, one might 
still} expect that the following part of the 
evolution is going to be quite boring since only a lower scale energy
transform will happen until the rest of the energy of the exciting pulse
disperse and the system settles down to the static monopole.

Contrary to this simple hypothetical scenario quite interesting
features of the underlying non-linear system show up. First of all, it
was found that the exciting pulse leaves the central region so that it
drags some part of the energy of the original static monopole  with
itself. This process is justified by the fact that in the central
region, where a long-lasting quasi-stable 'breathing state' develops,
at least at the beginning of the evolution the average energy density
is always less than that of the static monopole was. Since the
original static magnetic monopole is stable we know that by the end
the missing energy has to come  back to the central region. However,
as our numerical simulations indicates \cite{fr2} this process is
unexpectedly slow and, in certain cases, behaves completely contrary
to our expectations. For instance, the difference between the
dynamical and the original static energies contained in a ball of
radius $r$ may be positive or negative  depending on the strength of
the non-linear aspects of the dynamics, as well as, on location,
i.e. on the value of $r$. In particular, the time dependence of this
``extra energy'' content can be characterized by its high frequency
oscillating part and by its mean value. We found a ``universal'' time
decay for the amplitude of the high frequency oscillations of the form
of a power law with exponent close to the ``ideal'' value
$-5/6$. However, the 
exponent of the similar power law time decay of the mean value varies
with the radius of the ball (see subsection \ref{et2}).

It is also a characteristic feature of the evolution that some part of
the energy of the exciting pulse is getting to be stored by
expanding shells of oscillations of the massive Yang-Mills field built
up in the distant region. These shells of oscillations behaves much
like the oscillations of a simple massive Klein-Gordon field (see \cite{fr1}
for a detailed investigation) since the coupling of the
Yang-Mills and Higgs variables is negligible there. The monopole can
pull back the missing energy, approaching the static
configuration in the central region, only before these shells get
too far away.

Due to the coupling of the Yang-Mills and Higgs fields in the central
region, along with the ``breathing'' of the monopole there, the
massless Higgs field continuously take away a small  fragment of the
energy of the monopole, which is radiated to future null
infinity. The corresponding decrease of the total energy which can be
associated with the hyperboloidal hypersurfaces decreases in time with
power $-2/3$ (see subsection \ref{et1}). Note, however, that this
total energy (it would be called as the `Bondi energy' in general
relativity), as well as, its limit value, is always strictly  larger
than the energy of the original static monopole.
This is because the expanding shells of oscillations store
some part of the energy of the original pulse forever. In fact, this
part of the energy cannot reach future null infinity since it is
associated with the massive Yang-Mills field, moreover, as they are
getting further and further away from the central region the monopole
can get back less and less energy from them since non-linear effects
are getting to be more and more  negligible. As it is shown in
subsection \ref{et2}, the energy stored by these expanding shells also
approaches its non-zero asymptotic value from above following a power
law time decay with exponent $-2/3$.  

The time evolution of the frequency of oscillations of the excited
monopole behaves also somewhat counter to the general
expectations. Consider first the fundamental frequency of the
``breathing'' monopole.  Instead of having a slow decay in the
frequency of the associated oscillations as their amplitude decreases
there is an increase in the frequency which takes the maximum value at
the `end of time', i.e. at future timelike infinity. Interestingly,
while the frequency of the breathing monopole tends to the mass of the
Yang-Mills field from below the frequency of the oscillations stored
in the expanding shell has no upper bound, although the amplitude is
decreasing, as in case of the massive Klein-Gordon field \cite{fr1}.
Note that these phenomena are not at all unphysical. Recall that a
simple physical system like the spinning coin on the desk can produce
a quite similar effect. As energy loss happens due to dissipative
processes the amplitude of the ``spinning'' is getting smaller and
smaller in consequence of which the frequency of the oscillations is
getting higher and higher. In case of the distant oscillating shells
the unbounded increase of the frequency is, in fact, to  compensate
the decrease of the oscillation amplitudes which together ensure the
conservation of the energy stored by the shells.

In virtue of these results it seems not to be over-rating to say that
the investigation of the dynamics properties of magnetic monopoles
made transparent a number of interesting and unexpected features of
the underlying non-linear system. Hopefully, these sort of
investigations will stimulate further numerical and analytic
investigations of various { similar} non-linear dynamical systems.
For instance, partly motivated by the findings of the above described
investigations, in a framework of linear  perturbation theory some of
the above mentioned features could be explained successfully both
qualitatively and quantitatively \cite{fv}. Note, however, that  truly
non-linear effects are by their very nature out of the scope of these
sort of investigations. { In particular,} as some of the new results of
this paper make it transparent, they cannot be used to describe the
behavior of the investigated system when the energy of the exciting
pulse is much larger than that of the original static monopole.

Although in this paper we use our numerical code to describe various
oscillations of a localized monopole, we would like to note that oscillations in
a nonlinear theory may be responsible for the very existence of some compact
objects. In a wide variety of nonlinear field theories, where no static soliton
solutions exist, almost periodic long living oscillon configurations have been
found.
See for example \cite{graham1,gleiser1,gleiser2,graham2,graham3,saffin} for
recent results on this quickly evolving topic. 
A slightly modified version of our numerical code have also been used in
\cite{ffpr} for a detailed study of oscillon configurations in $\phi^4$ scalar
theory. 

The structure of this paper is as follows. In the next section, after
recalling some of the basics related to the properties of the
underlying generic dynamical system a detailed description of the
specific choice for both  the YMH system and the geometry is
provided. The static hyperboloidal conformal gauge applied in our
investigations is introduced in section \ref{shcg}, while the first
order representation of the field equations, relevant for the used
conformal setting, is given in section \ref{fohs}.  A detailed
description of the applied numerical scheme is presented in section
\ref{scheme}, while various numerical tests of the code, for the case
of massive and massless Klein-Gordon field, are presented in section
\ref{tests}. Finally, all the numerical results concerning the
evolution of dynamical magnetic monopoles, including subsections
providing detailed description of { the time dependence of the
basic and derived field variables and that of the frequency of the
associated oscillations,} the energy transfers, monitoring of the 
numerical violation of constraints and the energy conservation, along
with the behavior of the radial and angular pressures, as well as,
the magnetic charge density, are presented in section \ref{numrez}.

\section{Preliminaries}\label{prelim}

The investigated dynamical magnetic monopole is described as a
coupled $SU(2)$  YMH system. The Yang-Mills field is represented by an
$\mathfrak{su}(2)$-valued vector potential $A_a$ and the associated 2-form
field $F_{ab}$ reads as
\begin{equation}
F_{ab}=\partial_aA_b-\partial_bA_a+i g\left[A_a,A_b\right]
\label{ymf}
\end{equation}
where $[\ ,\ ]$ denotes the product in $\mathfrak{su}(2)$ and $g$
stands for the gauge coupling constant. The Higgs field 
(in the
adjoint representation) 
is given by an $\mathfrak{su}(2)$-valued
function $\psi$ while its gauge covariant derivative reads as
$\mathcal{D}_a\psi= \partial_a\psi+i g[A_a,\psi]$. The dynamics of the
investigated YMH system is determined by the 
action
\begin{equation}
S=\int \left\{Tr(F_{ef}F^{ef})
+2\left[Tr(\mathcal{D}_e\psi\mathcal{D}^e\psi)-V(\psi)\right]
\right\}{\epsilon}, 
\label{act}
\end{equation}
where $\epsilon$ is the 4-dimensional volume element, moreover,
$V(\psi)$, describing the self-interaction of the Higgs field, is
chosen to be the standard quadratic potential
\begin{equation} \label{pot}
V\left(\psi\right)=\frac{\lambda}{4}\left[Tr(\psi^2)-H_0^2\right]^2,  
\end{equation}
where $\lambda$ and $H_0$ denote the Higgs self-coupling constant
and the `vacuum expectation value' of the Higgs field, respectively.   

The symmetric energy-momentum tensor of the considered YMH
system takes the form 
\begin{equation}\label{Tab}
T_{ab}=-\frac{1}{4\pi}\left[Tr(F_{ae}{F_b}^{e})
-Tr(\mathcal{D}_a\psi\mathcal{D}_b\psi)
+\frac{1}{4}g_{ab}\mathcal{L}\right],
\end{equation}
where $\mathcal{L}$ stands for the Lagrangian
\begin{equation}
\mathcal{L}=Tr(F_{ef}F^{ef})
+2\left[Tr(\mathcal{D}_e\psi\mathcal{D}^e\psi)-V(\psi)\right]\,.
\end{equation}
 
\subsection{Fixing the geometrical and gauge setup}\label{smooth}

This paper is to investigate the evolution of spherically symmetric
Yang-Mills--Higgs 
(YMH) systems on a flat Minkowski background spacetime. Accordingly,
as a fixed background, the four-dimensional 
Minkowski spacetime $(\mathbb{R}^4,\eta_{ab})$ will be applied, the
line element of which in the conventionally used Descartes-type
coordinates  $(x^0,x^1,x^2,x^3)$  reads as
\begin{equation}
ds^2=(dx^0)^2-(dx^1)^2-(dx^2)^2-(dx^3)^2.
\end{equation}

The gauge group is specified by giving the set of generators
$\{\tau_{_{I}}\}$ (I=1,2,3)  of the associated $\mathfrak{su}(2)$ Lie
algebra which reads as
\begin{equation}
\tau_{_{I}}=\frac{1}{2}\sigma_{_{I}},
\end{equation}
where $\sigma_{_{I}}$ denote the Pauli matrices.  The commutation
relations relevant for this choice of generators are  
\begin{equation}
\left[\tau_{_{I}},\tau_{_{J}}\right]=i {\varepsilon}_{_{IJK}}
\tau_{_{K}},
\end{equation}
where ${\varepsilon}_{_{IJK}}$ denotes the completely antisymmetric
tensor with ${\varepsilon}_{_{123}}=1$. 

In addition to the above special choice concerning the gauge group,
our considerations will be restricted to YMH systems which are yielded by
the `minimal' dynamical generalization of the static  t'Hooft-Polyakov
magnetic monopole configurations \cite{tH,P} (see also
\cite{Goddard,Hua}). Accordingly, the Yang-Mills and Higgs field
variables,  $A_a=A_a^{_{I}}\tau_{_{I}}$ and
$\psi=\psi^{_{I}}\tau_{_{I}}$, are specified, in the Coulomb gauge, via
the relations
\begin{equation}
A_{_0}^{_{J}}=0,\ \
A_{_I}^{_{J}}=\frac{(1-w)}{g} \varepsilon_{_{IJK}}
\frac{x^{_{K}}}{r^2} \label{rA1}
\end{equation}
\begin{equation}
\psi^{_{I}}=H\frac{x^{_{I}}}{r},\label{rP1} 
\end{equation}
where the functions $w$ and $H$ are assumed to be smooth functions in
$\mathbb{R}^4$ depending upon the coordinates $x^0,x^1,x^2,x^3$ only
in the combinations  $t=x^0$ and
$r=\sqrt{(x^1)^2+(x^2)^2+(x^3)^2}$. 

Since the metric, as well as the matter fields, are required to be
spherically symmetric, the use of the standard coordinates
$(t,r,\theta,\phi)$, adapted to the spherical symmetry of
$(\mathbb{R}^4,\eta_{ab})$, is the most suitable. In these coordinates 
the line element of the Minkowski metric takes the form
\begin{equation}
ds^2=dt^2-dr^2-r^2\left(d\theta^2+\sin^2\theta\, d\phi^2\right).\label{le}
\end{equation}

We have assumed above that $w$ and $H$ are smooth functions of $t$ and
$r$. This might be surprising especially because our eventual aim is to
carry out numerical simulations of the YMH systems under
considerations. Numerical methods are inherently too rough to make a
sensible distinction between configurations belonging to different
differentiability classes. In fact, our smoothness assumption
is to ensure certain technical conveniences used later and
it is supported by the following considerations. (Note that for the
following argument it would be enough to assume that $w$ and $H$ are
smooth in a sufficiently small neighborhood of the 
origin.) In numerical simulation
of spherically symmetric configurations a grid boundary representing
the origin inevitably appears. This means that we need to solve an
initial-boundary value problem which, in particular, requires the
specification of `boundary behavior' of $w$ and $H$ at the origin
throughout the time evolution. This is, in fact, the very point where
we make use of our smoothness assumption. In virtue of
(\ref{rA1}) and (\ref{rP1}) it is straightforward to see that
spherical symmetry, along with the required smoothness of $w$ and $H$
through the origin, ensures that in a neighborhood of the origin $w$ and
$H$ are even and odd functions of the $r$-coordinate,
respectively. Consequently, as it is described in
subsection\,\ref{num} in more detail, by extending our grid by a
suitable number of virtual grid points the `boundary values', for
instance, of the r-derivatives of $w$ and $H$ need not be specified by
hand, instead they are naturally yielded by the time evolution of the
corresponding odd and even functions.

The Yang-Mills field $A_a$ as it is given by (\ref{rA1}) is in the ``Coulomb
gauge''. Another frequently used gauge representation is the so called
``abelian gauge'' in which $\psi$ has only one non-vanishing component,
and which is
achieved by making use of the gauge transformation
\begin{equation}
U=\exp(i\theta \tau_{_{2}})\exp(i\phi \tau_{_{3}}).
\end{equation}
A straightforward calculation yields that in the
corresponding abelian gauge the Yang-Mills  and the Higgs fields read
as 
\begin{equation}
A_a=-\frac{1}{g}\left[w\left\{\tau_{_{2}}(d\theta)_a-
\tau_{_{1}}\sin\theta(d\phi)_a \right\} +
\tau_{_{3}}\cos\theta(d\phi)_a\right] 
\end{equation} 
and
\begin{equation}
\psi=H \tau_{_{3}}.  
\end{equation} 

The substitution of these gauge representations into the equations of
motion, deducible from the action (\ref{act}), yields the evolution
equations for $w$ and $H$ which are given as

\begin{eqnarray}
& & r^{2}{\partial ^{2}_r}{w}- r^{2}{\partial ^{2}_t}{w}
= w\left[\left({w}^{2}-1\right)+g^2{r^{2}}{H}^2 \right]
\label{ymhe22}  \\ 
& &r^{2}{\partial ^{2}_r}{H}+ 2r{\partial_r }{H}
- r^{2}{\partial ^{2}_t}H = \nonumber\\ & & 
\phantom{{\partial ^{2}_r}{H}+ 2r{\partial_r }{H}
}
H\left[2{w}^{2}+\frac{\lambda}{2}{r^{2}}(H^2-H_0^2) 
\right].  \label{ymhe11} 
\end{eqnarray}
The static finite energy solutions of these equations are called
't\,Hooft-Polyakov magnetic monopoles.  

\subsection{Regularity at the origin and at infinity}\label{reg}

These equations, along with the former assumption concerning the
smoothness of the basic variables, ensure certain regularity of the
Yang-Mills and Higgs fields. In particular, it turned out that the
values of $w$ and $H$ are restricted at the origin
and at spacelike infinity throughout the evolution. At the origin (see
appendix A for more details) the relations
\begin{equation}
w(t,0)=1, \ \ {\rm and} \ \ H(t,0)=0 \label{reorig}
\end{equation}
have to hold to have finite energy density, measured by the static
observer with four velocity $u^a=(\partial/\partial t)^a$. 
In addition, whenever the smoothness of $w$ and $H$ is guaranteed in a
neighborhood of the origin we also have 
\begin{equation}
\partial_r^k w\vert_{r=0}=0\label{reorig1}
\end{equation}
for $k$ being odd, while 
\begin{equation}
\partial_r^k H\vert_{r=0}=0\label{reorig2}
\end{equation}
for $k$ being even. For the first sight it might be a bit of surprise
but a same type of reasoning that lead to (\ref{reorig1}) and
(\ref{reorig2}) provides a restriction also for the rest of the
derivatives. Namely, from the field equations and from the smoothness
requirement it follows that at the origin all the non-vanishing
derivatives  $\partial_r^k w$, with $k\geq 4$, and $\partial_r^l H$,
with $l\geq 3$, can be given as a function  of $\partial_r^2 w$ and
$\partial_r H$, along with their various time derivatives.

The argument associated with the regularity of the basic variables at
spacelike infinity is more delicate. To get it we need to refer to the
conformal (non-physical) setting, where our  numerical simulation
actually will be carried out. In this framework it seems to be
essential to assume that the fields are at least $C^2$ even through
future null infinity, $\scrip$. We refer here to this requirement as
being merely a technical assumption, nevertheless, we would like to
emphasize that the use of it is  supported by the following
observations. In \cite{wini} it was proved that the time evolution of a
massive 
Klein-Gordon 
field in Minkowski spacetime with initial data of  compact support
necessarily yields 
${\mathcal{O}} (1/r^\infty)$ asymptotic behavior at null infinity. In
addition, the results of Eardley and Moncrief \cite{EM1,EM2,monc,BM}
concerning the local and global existence of YMH fields in
4-dimensional 
Minkowski spacetime supports that the above technical assumption can 
be deduced from the field equations at least in case of suitably
chosen initial data specifications. 

This smoothness requirement, along with the relevant form of the
rescaled field equations, implies (see appendix A for more details)
that $w$ has to vanish while $H$  has to be constant along the null
geodesic generators of $\scrip$. The field equations also yield
further restriction on  this constant limit value $H_{_{\infty}}$ of
$H$ at $\scrip$. In particular, $H_{_\infty}$ must take the value
$H_0$ whenever  $\lambda\not=0$ but it is an arbitrary (positive)
constant, $C$, otherwise. Thus, by making use of a limiting argument,
the values of $w$ and $H$ at spacelike infinity can be seen to be
determined as
\begin{eqnarray}
w_{_\infty}&=&\lim_{r \rightarrow \infty}w(t,r)=0, \\ 
H\hspace{-.08cm}{}_{_\infty}&=&\lim_{r \rightarrow \infty}H(t,r)=\left\{ 
\begin{array} {r@{,\quad}l} 
 H_0& if \quad \lambda\not=0;\\ C& otherwise.
\end{array} 
\right. \label{inf}
\end{eqnarray}
Notice that the later
relations are in accordance with our general expectations that the
considered dynamical YMH systems do really possess the asymptotic
fall-off properties of a magnetic monopole \cite{Goddard,Hua}
throughout the 
time evolution. 

It follows from equation (\ref{ymhe22}), (\ref{ymhe11}) and
(\ref{inf}) that the semi-classical vector boson and Higgs mass,
$M_{_w}$ and $M_{_H}$, are 
\begin{eqnarray}
M_{_w}&=&gH\hspace{-.08cm}{}_{_\infty}\\
M_{_H}&=&\sqrt{\lambda}H\hspace{-.08cm}{}_{_\infty}.
\end{eqnarray}
In particular, both fields are massless whenever
$H\hspace{-.08cm}{}_{_\infty}$ vanishes, while only the Higgs field is
massless, although it is coupled to a massive vector boson, 
whenever the self coupling constant $\lambda$ is zero. 

Whenever the smoothness of
the basic variables is guaranteed through $\scrip$ then the $n^{th}$
order $r$-derivatives of $w$ and $H$ are also restricted there. In
particular, it follows from the relations (\ref{ctr}),
(\ref{resinf})-(\ref{resinf2}), along with the vanishing of the
energy-momentum expressions at null infinity that whenever
$H\hspace{-.08cm}{}_{_\infty}\not=0$ the asymptotic fall off
conditions, 
\begin{equation}
\lim_{r \rightarrow \infty}{\left[r^{2n}\partial^n _r  w(t,r)\right]}
=0 \ \ ({\rm for}\ n\geq0)\label{inf2}
\end{equation}
and
\begin{equation}
\lim_{r \rightarrow \infty}{\left[r^{2n}\partial^n _r  H(t,r)\right]}
= 0, \ \ ({\rm for}\ n\geq 2)\label{inf22}
\end{equation}
have to be satisfied.

It is important to note that whenever $\lambda=0$, i.e. the
self-coupling of the Higgs field is turned down, there exists an
explicitly known \cite{ps} static solution to (\ref{ymhe22}) and
(\ref{ymhe11})  
\begin{eqnarray}
w_s&=& \frac{g C r}{{\rm sinh}(g C r)}\label{ws}\\
H_s&=& C\left(\frac{1}{{\rm tanh}(g C r)}-\frac{
1}{g C r}\right),\label{hs}
\end{eqnarray}
where $C$ is an arbitrary positive constant which is in fact the limit
value, $H\hspace{-.08cm}{}_{_\infty}$, of the r.h.s. of
(\ref{hs}). This static solution is called to be the
Bogomolny-Prasad-Sommerfield  (BPS) magnetic monopole which is known
to be a linearly stable configuration \cite{baa}. 

Notice that in the BPS limit, i.e. whenever $\lambda=0$ and
$H\hspace{-.08cm}{}_{_\infty}=H_0\not= 0$, the Higgs field becomes
massless and the only scale parameter of the system is the vector
boson mass $M_w=gH\hspace{-.08cm}{}_{_\infty}$.  Since in the case
considered here $H\hspace{-.08cm}{}_{_\infty}\not=0$, the re-scalings
$t\rightarrow \tilde t=t M_w$, $r \rightarrow \tilde r=r M_w$ and
$H\rightarrow \tilde H=H/H\hspace{-.08cm}{}_{_\infty}$ transform the
parameters to the value  $g=H\hspace{-.08cm}{}_{_\infty}=1$.  This
implies that whenever we would like to study the excitations of the
BPS monopole it suffices to consider the time evolution of the  system
only for the particular choice of the parameters
$g=H\hspace{-.08cm}{}_{_\infty}=1$,  because by inverse re-scalings of
these solutions all the possible solutions to the field equations can
be generated. Actually, to check the efficiency of our numerical
implementation of the evolution equations first the stability of this
static solution was investigated. Later, the complete non-linear
evolution of systems yielded by strong impulse type excitations of
these analytic static solution was also studied in detail.

Although it is not necessary to be done, nevertheless we preferred to
put the principal part of (\ref{ymhe11}) into the same form as the
principal part of (\ref{ymhe22}) which was achieved by making use of
the substitution
\begin{equation}
H(t,r)=\frac{h(t,r)}{r}+H\hspace{-.08cm}{}_{_\infty}.\label{h}
\end{equation}
It follows then that our new basic variable $h$ vanishes at the origin
while it takes a finite limit value at infinity if $H$ is
guaranteed to tend to its boundary value fast enough there. The
substitution of (\ref{h}) into the relations (\ref{ymhe22}) and
(\ref{ymhe11}) yields then  
\begin{eqnarray}
&&\hspace{-.3cm}r^{2}\left({\partial ^{2}_r}{w}-{\partial ^{2}_t}{w}\right)=
w\left[\left({w}^{2}-1\right)+g^2(h+H\hspace{-.08cm}{}_{_\infty}
r)^2\right]\phantom {H+} \label{ymhe222}\\ 
&&\hspace{-.3cm}r^{2}\left({\partial ^{2}_r}{h}-{\partial
  ^{2}_t}h\right)= \nonumber \\ &&\hspace{1.2cm}
(h+H\hspace{-.08cm}{}_{_\infty} r) 
\cdot\left[2{w}^{2}+
\frac{\lambda}{2}(h^2+2H\hspace{-.08cm}{}_{_\infty} r h) 
\right], \label{ymhe111}
\end{eqnarray}
where we have used the relation $H\hspace{-.08cm}{}_{_\infty}=H_0$
whenever $\lambda\not=0$. The regularity of the solutions to these
equations at $r=0$ 
follows from the boundary conditions (\ref{reorig}). It is important
to note that (\ref{inf}) and (\ref{h}), along  with our smoothness and
symmetry assumptions, which  in particular implies that $H$ is an odd
function of $r$, guarantees that $h+H\hspace{-.08cm}{}_{_\infty} r$
has to be an even function of the $r$-coordinate. Moreover, since
$h^2+2H\hspace{-.08cm}{}_{_\infty} r h=(h+H\hspace{-.08cm}{}_{_\infty}
r)^2-{H^2\hspace{-.08cm}{}_{_\infty}} r^2$ the right hand  sides of
(\ref{ymhe222}) and 
(\ref{ymhe111}) are both even functions of the $r$-coordinate. Notice
finally that, by equations (\ref{ymhe222}) and (\ref{ymhe111}), the
expressions ${\partial ^{2}_r}{w}-{\partial ^{2}_t}{w}$ and ${\partial
^{2}_r}{h}-{\partial ^{2}_t}h$ have regular limits at $r=\infty$
because, in virtue of (\ref{inf}),(\ref{inf2}),(\ref{inf22}) and
(\ref{h}), $w$ and $h$ remain finite while $r$ tends to infinity. 

\section{The static hyperboloidal conformal gauge}\label{shcg}

In this section a conformal transformation of the
Minkowski spacetime -- along with the relevant form of the above
matter field equations -- will be considered. 

The applied conformal transformation is a slightly modified version of
the 
gauge transformation applied first by Moncrief \cite{mon} (see also
\cite{hu,zs}). It is defined by introducing the new coordinates $(T,R)$
instead of $(t,r)$ as
\begin{eqnarray}
T(t,r)&=& \kappa t-\sqrt{\kappa^2 r^2+1} \label{TR21}\\
R(r)&=&\frac{\sqrt{\kappa^2 r^2+1}-1}{\kappa r} \label{TR22}   
\end{eqnarray} 
with the inverse relations
\begin{equation}
t=\frac{1}{\kappa}\left(T+\frac{1+R^2}{1-R^2}\right)\ \ {\rm
  and} \ \   r=\frac{2R}{\kappa(1-R^2)}, \label{iTR2}  
\end{equation}
where $\kappa$ is an arbitrary positive constant.  The Minkowski
spacetime is covered by the coordinate domain  satisfying the
inequalities  $-\infty < T < +\infty $ and $0 \leq R < 1$ (see
Fig.\ref{fM2}).

The line element of the conformally rescaled metric $\widetilde{g}_{ab}
=\Omega^2{g}_{ab}$ in coordinates $(T,R,\theta,\phi)$ takes the form
\begin{eqnarray}
&&\hspace{-.3cm}d\widetilde
  s^2=\frac{\Omega^2}{\kappa^2}dT^2+2RdTdR-dR^2\nonumber\\ 
&&\hspace{-.3cm}\phantom{d\widetilde
    s^2=\frac{\Omega^2}{\kappa^2}dT^2++}-R^2\left(d\theta^2+{\rm
    sin}^2\theta\,d\phi^2\right), 
\end{eqnarray}
where the conformal factor is
\begin{equation}
\Omega(R)=\frac{\kappa}{2}(1-R^2),\label{Om2}
\end{equation}
and, by (\ref{iTR2}) and (\ref{Om2}), we have the
relation
\begin{equation}
r\Omega=R.\label{rOm2}
\end{equation}

In this conformal representation the $R=1$ coordinate line represents
$\scri^+$ through which the metric $\widetilde{g}_{ab}$ smoothly
extends to the coordinate domain with $R\geq 1$.

The name `static hyperboloidal gauge' is explained by the
following observations. First, (\ref{TR22}) tells us that the $R=const$
lines represent world-lines of `static observers', i.e. integral
curves of the vector field $(\partial/ \partial t)^a$. Second, it
follows from  (\ref{TR21}) that the $T=const$ hypersurfaces are, in
fact, hyperboloids satisfying the relation
$(\kappa t-T)^2-\kappa^2 r^2=1$ in the Minkowski spacetime.

The equation describing radial null geodesics in the $(T,R)$ coordinate 
system is independent of the parameter $\kappa$,
\begin{equation}
T=\frac{\pm2R}{1\pm R}+T_0 \ , \label{nullgeod}
\end{equation}
with the plus signs corresponding to outgoing and the minus signs to
ingoing geodesics. 
The form of the constant is chosen in a way that both type of 
geodesics cross the origin $R=0$ at time $T=T_0$.
\begin{figure}[ht]
\unitlength1cm
\centerline{
\begin{minipage}[t]{8.cm}
 \centerline{
  \epsfysize=6.cm
\epsfbox{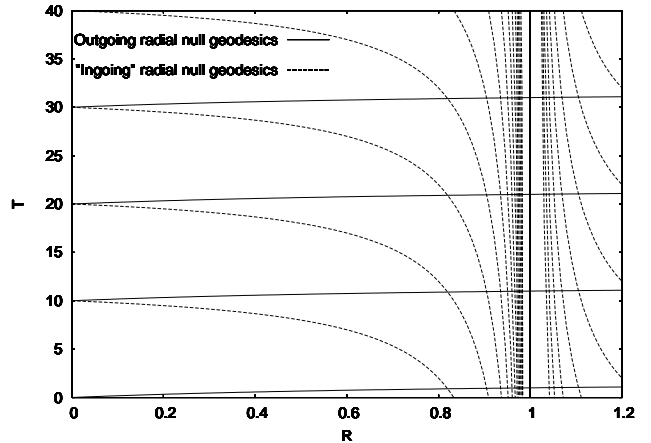}
 }
\caption{\footnotesize \label{fMM}   { The ingoing and outgoing radial
null geodesic curves are shown in both the ``physical'' and
``unphysical'' regions. As in other conformal representations each
point represents a two-sphere of radius ${R}$. The interior of the
strip $0\leq R<1$  corresponds to the original Minkowski spacetime
while $\scri^+$ is represented by the $R=1$ coordinate line. Notice
that the $R=const$ lines possesses timelike character everywhere
except at the $R=1$ line which is null.}}
\end{minipage}
}
\end{figure}
The outgoing geodesic emanating from the origin reaches future null
infinity $R=1$ in a finite coordinate time at $T=T_0+1$.  The ingoing
geodesics starting from a point at radius $R_0$ at $T=0$ reaches the
origin at $T=2R_0/(1-R_0)$.  Consider now a geodesic starting at a
point close to $\scrip$ with radius $R_0=1-\delta R$.  Such a geodesic
reaches the origin at $T=-2+2/\delta R$.  This means that for
geodesics coming in from the far away region, the travel time
essentially doubles when the geodesic starts from  a point ``twice as
close'' to the $R=1$ line.  This has an important consequence in
relation to the validity  domain of our numerical code.  Namely, the
doubling of the resolution yields the doubling of the time interval
within which we may expect our numerical simulation to provide a
proper solution to the selected problem.  Here it is assumed tacitly
that all the possible inaccuracies born at and coming from the
outer region -- that is the part of the spacetime where the evolution
cannot be described properly by any of the numerical techniques based
on a finite grid -- travel  inwards only with the speed of light.

Consider now the field equations relevant for this conformal
setting. To start off take a function $f=f(t,r)$ of the
coordinates $t$ and $r$ and denote by $\widetilde f=\widetilde f(T,R)$
the function 
$f(t(T,R),r(T,R))$ yielded by the substitution of (\ref{iTR2}) into
$f$. A straightforward  calculation justifies then that for any pair of
functions $f=f(t,r)$ and $\widetilde f=\widetilde f(T,R)$ 
the relation
\begin{eqnarray}
&& \hskip-.5cm r^2\left({\partial ^{2}_r}{f}-{\partial ^{2}_t}f\right)=
\frac{4R^2}{(R^2+1)^2}\left(\frac{\Omega^2}{\kappa^2}{\partial ^{2}_R}
{\widetilde f}-{\partial^{2}_T}{\widetilde
f} \right.\label{tr2} \\ &&
\hskip-.4cm\left.-2R{\partial_R\partial_T}{\widetilde f}
-\frac{2\Omega}{\kappa(R^2+1)}{\partial_T}{\widetilde
f}-\frac{\Omega R(R^2+3)}{\kappa(R^2+1)}{\partial_R}{\widetilde
  f}\right) \nonumber
\end{eqnarray}
holds. It follows then from (\ref{rOm2}) and (\ref{tr2}) that the field
equations, (\ref{ymhe222}) and (\ref{ymhe111}), in the current
conformal representation read as
\begin{eqnarray}
& & \hskip-.95cm \frac{4R^2}{(R^2+1)^2}{\mathfrak P}{\widetilde w}
= \widetilde w\hskip-.07cm\left[\hskip-.07cm\left({\widetilde
    w}^{2}-1\right)\hskip-.07cm+\hskip-.07cm 
g^2\hskip-.07cm\left(\widetilde h+ H\hspace{-.08cm}{}_{_\infty}
R\Omega^{-1}\right)\hskip-.07cm^2\hskip-.07cm\right] \label{2ymhe24}\\
& & \hskip-.95cm\frac{4R^2}{(R^2+1)^2}{\mathfrak P}{\widetilde h}=
\left({\widetilde h}+H\hspace{-.08cm}{}_{_\infty} R\Omega^{-1}\right)
\nonumber\\
& & \hskip-.9cm\phantom{\hskip-.8cm\frac{4R^2}{(R^2+1)^2}{\mathfrak
    P}= \left( \right) \frac{4R^2}{(R^2)}}
\cdot \left[2{\widetilde w}^{2}+\frac{\lambda}{2}\widetilde
  h\left(\widetilde 
h+2H\hspace{-.08cm}{}_{_\infty} R\Omega^{-1}\right)
\right]\hskip-.12cm, \label{2ymhe14} 
\end{eqnarray}
where the second order partial differential operator $\mathfrak P$ is
defined as 
\begin{eqnarray} \label{difop}
&&{\mathfrak P}=\frac{\Omega^2}{\kappa^2}{\partial ^{2}_R} - {\partial
  ^{2}_T}- 
2R{\partial_R\partial_T}-\frac{2\Omega}{\kappa(R^2+1)}{\partial_T}
  \nonumber\\&&\phantom{{\mathfrak P}=
  \frac{\Omega^2}{\kappa^2}{\partial ^{2}_R} - {\partial 
  ^{2}_T}-}-
\frac{\Omega R(R^2+3)}{\kappa(R^2+1)}
{\partial_R}.
\end{eqnarray}

As before, the regularity of the solutions to these equations at the
origin, i.e. at 
$R=0$ is guaranteed by the regularity of (\ref{ymhe222}) and
(\ref{ymhe111}) at $r=0$, while, the regularity at future
null infinity, i.e. at $R=1$, follows from (\ref{inf}),
(\ref{inf2}) and (\ref{inf22}) provided that the fields have at least
$C^2$ extensions through $\scri^+$.

Hereafter, unless indicated otherwise, all the functions will be
assumed to depend only on the coordinates $T$ and $R$. Thereby we 
suppress all of the {\it tildes}, introduced above. 

\section{The first order hyperbolic systems}\label{fohs}

This section is to derive a first order hyperbolic system
from the above evolution equations for which the initial value problem
is well-posed. To see that equations  (\ref{2ymhe24}) and
(\ref{2ymhe14}) can be put into the form of
a strongly hyperbolic system we shall follow a standard process (see
e.g. \cite{ch}). Correspondingly first we introduce the first
order derivatives of $w$ and $h$,   
\begin{equation}
w_T= {\partial_T w}, \ \ 
w_R= {\partial_R {w}}, \ \  
{ h}_T= {\partial_T { h}}, \ \  
{ h}_R= {\partial_R { h}},\label{nw}
\end{equation}
as new variables. In terms of the relevant enlarged set of
dependent variables 
(\ref{2ymhe24}) and (\ref{2ymhe14}) can be given as 
\begin{eqnarray}
{\partial _T}{ w}_T&=& 
\frac{\Omega^2}{\kappa^2} \left({\partial_R}{
w}_R\right)-2R\left({\partial_R}{w}_T\right)
        +
b_{ w} \label{2ymhe25}\\  
{\partial _T}{ h}_T&=& \frac{\Omega^2}{\kappa^2} \left({\partial_R}{
h}_R\right)-2R\left({\partial_R}{h}_T\right) 
        +
b_{ h}, \label{2ymhe15}
\end{eqnarray}
where $b_{ w}$ and $b_{ h}$ are given as 
\begin{eqnarray}
& &\hskip-.3cm b_{ w}= -\frac{2\Omega}{\kappa(R^2+1)}{ w}_T-\frac{\Omega
R(R^2+3)}{\kappa(R^2+1)}{ w}_R \label{bw2}\\& & \hskip.3cm
  -\frac{(R^2+1)^2}{4\Omega^2 R^{2}} 
w\left[\Omega^2\left({ w}^{2}-1\right)\hskip-.07cm+ g^2
\left(\Omega h+ H\hspace{-.08cm}{}_{_\infty} R\right)^2\right]\hskip-.07cm,\nonumber\\
& &\hskip-.3cm  b_{ h}=-\frac{2\Omega}{\kappa(R^2+1)}{ h}_T-\frac{\Omega
R(R^2+3)}{\kappa(R^2+1)}{ h}_R \nonumber\\& & \hskip1.4cm
-\frac{(R^2+1)^2}{4\Omega^2R^2}  
\left(\Omega{ h}+H\hspace{-.08cm}{}_{_\infty} R\right) \label{bh2}\\& &
\hskip2.4cm  
\cdot\left[2\Omega{ w}^{2}+\frac{\lambda}{2} h\left(\Omega
h+2H\hspace{-.08cm}{}_{_\infty} R\right) \right].\nonumber
\end{eqnarray}

It is straightforward to see that the above equations,
along with the 
first and third equations of (\ref{nw}), and the integrability
conditions 
\begin{equation}
{\partial_T}{ w}_R= {\partial_R} { w}_T\ \ \ {\rm
and}\ \ \ {\partial_T}{ h}_R= {\partial_R} { h}_T,
\end{equation}
possess the form of a first order system
\begin{equation}
{\partial_T}\Phi=A_R({\partial_R}\Phi)+B,\label{hyp}
\end{equation}
where $\Phi$ and $B$ are given as
\begin{equation}
\Phi=\left(
\begin{array}{c}
{ w} \\
{ w}_T \\
{ w}_R \\
{ h} \\
{ h}_T \\
{ h}_R \\
\end{array}
\right),\ \ 
B=\left(
\begin{array}{c}
{ w}_T \\
b_{ w} \\
0 \\
{ h}_T \\
b_{ h} \\
0 \\
\end{array}
\right),\ \ 
  \label{coeff}
\end{equation}
moreover, $A_R$ takes the form 
\begin{equation}\label{AR}
A_R= \left(
\begin{array}{cccccc}
0 & 0 & 0 & 0 & 0 & 0 \\
0 & -2R & {\Omega^2}/{\kappa^2}  & 0 & 0 & 0 \\
0 & 1 & 0 & 0 & 0 & 0 \\
0 & 0 & 0 & 0 & 0 & 0 \\
0 & 0 & 0 & 0 & -2R & {\Omega^2}/{\kappa^2} \\
0 & 0 & 0 & 0 & 1 & 0 \\
\end{array}
\right).
\end{equation}
Since the eigenvectors of $A_R$ comprise a complete system and its
eigenvalues are all real this first order system is, in fact, a
strongly hyperbolic system \cite{gko}. 

Notice also that $B$ does not depend on the spacelike derivatives of
$\Phi$, i.e. it is a functional of $T$, $R$ and $\Phi$ exclusively,
$B=B(T,R;\Phi)$. 
It is also straightforward to verify that the constraint equations, i.e.
the second and the fourth equations of (\ref{nw}), are preserved by the
time evolution governed by (\ref{hyp}). This, in particular,
guarantees that the pair of functions $w$ and $h$
yielded by the evolution of suitable initial data specifications
will automatically satisfy the original evolution equations
(\ref{2ymhe24}) and (\ref{2ymhe14}) as well,  provided that the
constraint equations hold on the initial data surface.

Finally, we would like to emphasize that the system specified by the
relations (\ref{hyp}) - (\ref{AR}) is not
only strongly hyperbolic but can be put into the form of first order
symmetric hyperbolic system. To see this introduce the variables $\hat
w_R$ and $\hat h_R$ by the relations 
\begin{equation}
\hat w_R=\frac{\Omega}{\kappa}w_R \ \ \ {\rm and}\ \ \ \hat
h_R=\frac{\Omega}{\kappa}h_R. 
\end{equation} 
Then the system of field equations for the vector variable
$\Phi=(w,w_T,\hat w_R,h,h_T,\hat h_R)^T$, possess the form of
(\ref{hyp}) with  
\begin{equation}
A_R= \left(
\begin{array}{cccccc}
0 & 0 & 0 & 0 & 0 & 0 \\
0 & -2R & {\Omega}/{\kappa}  & 0 & 0 & 0 \\
0 & {\Omega}/{\kappa} & 0 & 0 & 0 & 0 \\
0 & 0 & 0 & 0 & 0 & 0 \\
0 & 0 & 0 & 0 & -2R & {\Omega}/{\kappa} \\
0 & 0 & 0 & 0 & {\Omega}/{\kappa} & 0 \\
\end{array}
\right).
\end{equation} 
Nevertheless, in all of our numerical simulations the strongly
hyperbolic form given by (\ref{hyp}) - (\ref{AR}) was used which
system was found to be as  efficient in accuracy in a number of
particular cases as the symmetric hyperbolic system.

\section{The numerical scheme}\label{scheme}

\begin{figure}[ht]
\unitlength1cm
\centerline{
\begin{minipage}[t]{8.cm}
 \centerline{
  \epsfysize=6.5cm 
  \epsfbox{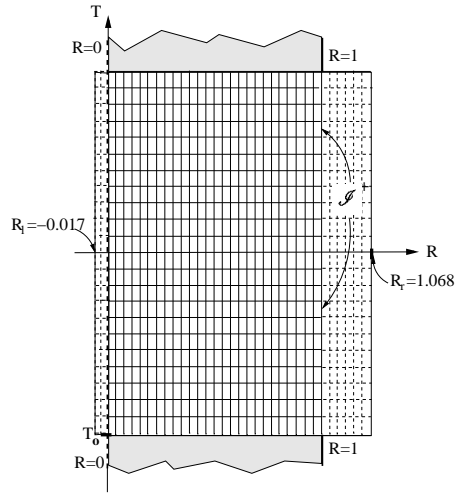}
 }
\caption{\footnotesize \label{fM2}   {The domains above the initial
data hypersurface $T=T_0$ covered by the simplest possible
orthogonal grid are indicated in the applied 
conformal representations. }} 
\end{minipage}
}
\end{figure}

In our numerical simulations we shall use the simplest possible
orthogonal grid based on the $T=const$ and $R=const$ 'lines' in the
domain $T>T_0$ as it is indicated on Fig.\ref{fM2}. 
The relevant discrete set of grid points is given as 
\begin{eqnarray}
T_l&=&T_0+l\Delta T, \ \ \  \ l=0,1,...,L_{max}\\
R_i&=&i\Delta R, \ \ \  \  \ \ \ \ \ \ i=0,1,...,I_{max},
\end{eqnarray}
for some $I_{max},L_{max}\in \mathbb{N}$ fixed numbers and with the relation
$\Delta T=k\Delta R$ for some $k\in \mathbb{N}$. In this setting a
function $f=f(T,R)$ will be represented by its values at the indicated
grid points, i.e. by $f^l_i=f(T_l,R_i)$. 

In our numerical simulations the total number of spatial grid points
was chosen to be an integer power of 2, i.e. $2^N$, for some value
$N\in \mathbb{N}$. A small fragment of these spatial gridpoints were
used to handle the grid boundaries at the origin and beyond future
null infinity (see subsection \ref{num} for more retails). For this
purpose we always used $1/2^6$ and  $1/2^4$ parts of the spatial
gridpoints on the left side of the origin and beyond future null
infinity. This means that the formal values of the $R$-coordinate at
these timelike portions of the numerical grid boundaries are
$R_l=-\frac{4}{256-4-16}\sim -0.017$ and
$R_r=1+\frac{16}{256-4-16}\sim 1.068$. Note that the appropriate
treatment of the grid boundaries does not require the use of fixed
portions of the spatial gridpoints. Nevertheless, we used this simple
approach to have a straightforward  setting in which we could compare
the results of our numerical simulations for various resolutions for
the price that only the $236/256$ part of the spatial gridpoints
represented points from the original Minkowski spacetime.

\subsection{The time integrator}

The time integration of (\ref{hyp}) is based on the use of the `method
of lines' in a higher order scheme as it is proposed by Gustafsson
{\it et al} \cite{gko}. In particular, we integrate (\ref{hyp}), along the
constant $R_i$ lines, by making use of a fourth order Runge-Kutta
scheme. This is done exactly in the manner as the Runge-Kutta
scheme is used to integrate first order ordinary differential
equations. Correspondingly, the value of $\Phi$ after a `time step' is
determined as   
\begin{equation}
\hspace{-.12cm}\Phi^{l+1}_i={\Phi^{l}}_i
+\frac{1}{6}\bigl({\Psi_{_{(I)}}}^{l}_i + 
2{\Psi_{_{(II)}}}^l_i+ 2{\Psi_{_{(III)}}}^l_i
+{\Psi_{_{(IV)}}}^l_i\bigr),  
\end{equation}
where
\begin{eqnarray}
&&\hspace{-.9cm}{\Psi_{_{(I)}}}^{l}_i= \Delta
  T\Bigl[A_R\bigl({\partial_R}\Phi^{l}_i\bigr)+ 
B\bigl(T_l,R_i;\Phi^{l}_i\bigr) \Bigr], \\
&&\hspace{-.9cm}{\Psi_{_{(II)}}}^{l}_i=\Delta T \Bigl[A_R\bigl({\partial_R}
\bigl[\Phi^{l}_i+\frac{1}{2}{\Psi_{_{(I)}}}^{l}_i\bigr]\bigr)
\nonumber\\& & \hspace{+1.2cm}+    
B\bigl(T_l+\frac{1}{2}\Delta T,R_i;\Phi^{l}_i+\frac{1}{2}
{\Psi_{_{(I)}}}^{l}_i\bigr) \Bigr], \\
&&\hspace{-.9cm}{\Psi_{_{(III)}}}^{l}_i=\Delta T \Bigl[A_R\bigl({\partial_R}
\bigl[\Phi^{l}_i+\frac{1}{2}{\Psi_{_{(II)}}}^{l}_i\bigr]\bigr)
\nonumber\\& & \hspace{+1.2cm}+    
B\bigl(T_l+\frac{1}{2}\Delta T,R_i;\Phi^{l}_i+\frac{1}{2}
{\Psi_{_{(II)}}}^{l}_i\bigr) \Bigr]\hspace{-.03cm}, \\
&&\hspace{-.9cm}{\Psi_{_{(IV)}}}^{l}_i=\Delta T \Bigl[A_R\bigl({\partial_R}
\bigl[\Phi^{l}_i+{\Psi_{_{(III)}}}^{l}_i\bigr]\bigr) \nonumber\\&
&\hspace{+1.2cm}+ 
B\bigl(T_l+{\Delta T},R_i;\Phi^{l}_i+
{\Psi_{_{(III)}}}^{l}_i\bigr) \Bigr],
\end{eqnarray}

To be able to apply this method of time integration we must calculate
$R$-derivatives of certain functions several times. In this
time integration process we approximated these $R$-derivatives by a
symmetric fourth order stencil (see appendix B). 

It follows from general considerations \cite{gko} that the above
described time integration process applied to 
(\ref{hyp}) cannot be stable unless a suitable dissipative term
is added to each evaluation of the right hand side of (\ref{hyp}). An
appropriate dissipative term, relevant for the fourth order Runge-Kutta
scheme used here, reads as   
\begin{equation}
\mathfrak{D}=\sigma (\Delta R)^5 \bigl({\partial^6_R}\Phi\bigr), 
\label{dissip}
\end{equation}
where $\sigma$ is a non-negative constant
and the sixth order $R$-derivatives were evaluated in a symmetric
sixth order stencil (see appendix B). 

Notice that the use of this dissipative term does not reduce the order
of accuracy of the applied finite difference approximation. Moreover, since
(\ref{hyp}) is an almost everywhere symmetrizable strongly hyperbolic
system and all the differential
operators are centered it follows from Theorem 6.7.2.\,of \cite{gko}
that the applied time integration process is stable provided $\sigma$
is sufficiently large and $k=\Delta T/\Delta R$ is sufficiently
small. Numerical experiments showed us that, for instance, the
particular choices $\sigma \sim 10^{-2}$ and $k=1/8$ provide the
required stability for the time integration of our evolution equation.
  
\subsection{The evaluation of the evolving fields at the origin and at
future null infinity}

Before being able to make the first time step, and later at each of
the time levels
we have to face the following technical problem. The source term $B$
in (\ref{hyp}) contains terms -- see the
last terms of (\ref{bw2}) and (\ref{bh2}) -- which are given as ratios
of expressions vanishing at the origin and at 
conformal infinity, respectively. The evaluation of these type of
expressions generally is a hard numerical problem.

It turned out, however, that the exact value of $b_{ w}$ and $b_{ h}$
is either irrelevant or can be shown to be zero. To see this consider
first the evolution of the fields at the origin. By making use of the
field equations and assuming that the fields are at least of class
$C^2$ at $R=0$ we have shown (see (\ref{reorig}) - (\ref{reorig2}))
that in
a sufficiently small neighborhood of $R=0$ the field values $w$ and
$h$ must possess the form 
\begin{eqnarray}
&& w=1+\hat w(T,R)\cdot R^2\\ 
&& h=\hat h(T,R)\cdot R^2-\frac{R}{\Omega} H\hspace{-.08cm}{}_{_\infty}, 
\end{eqnarray}
where $\hat w(T,R)$ and $\hat h(T,R)$ are assumed to be sufficiently
regular functions of their indicated variables. These relations
immediately imply that 
\begin{eqnarray}
&& \hspace{.24cm}w(T,0)\equiv 1,\\
&& w_T(T,0)\equiv 0,\\
&& w_R(T,0)\equiv 0,
\end{eqnarray}
moreover, 
\begin{eqnarray}
&& \hspace{.24cm}h(T,0)\equiv 0,\\
&& h_T(T,0)\equiv 0,\\
&& h_R(T,0)\equiv -\frac{2 H\hspace{-.08cm}{}_{_\infty}}{\kappa},
\end{eqnarray}
hold. In other words, the time evolution of our basic variables at the
origin is trivial and hence the evaluation of the $b_{ w}$ and $b_{ h}$
is not needed there. 

Despite their apparent singular behavior the evaluation of $b_{w}$
and $b_{h}$ at $R=1$ is possible based on the following
observations. The first two terms of $b_{w}$ and $b_{h}$ are
proportional to the conformal factor $\Omega$ so they vanish at
$\scrip$. The last terms also vanish there in spite of the presence of
the $\Omega^{-2}$ factors in them since these expressions always
contain as a  multiplying factor at least one of the massive field
variables, either $w$ or, whenever $\lambda\not=0$, $h$. These field
variables, however, possess the fall off property that $r^n w$ and
$r^n h$ for arbitrary positive integer value $(n\in\mathbb{N})$ tend
to zero while $r\rightarrow \infty$. By making use of this fact, along
with the relation $r=R/\Omega$, it is straightforward to check that
both $b_w$ and $b_h$ must vanish at $\scrip$. Accordingly, the values
of $b_w$ and $b_h$ were kept to be identically zero at $R=1$ in the
numerical simulations.  

\subsection{Increasing accuracy}

In order to ensure higher order of accuracy we also have applied the
following trick. Instead of calculating  the time evolution of the
variables $w$ and $h$ themselves we determined the evolution of the
deviation, $w_{_\Delta}=w-w_0$ and $h_{_\Delta} =h-h_0$, of them with
respect to certain analytic functions, $w_0$ and $h_0$. In fact, the
functions 
$w_0$ and $h_0$ need not necessarily required to be solutions of the
field equations although it is favorable to assume that they possess
the same type of  behavior at $R=0$ and at $\scri^+$ as the functions
$w$ and $h$ themselves.  This way it was possible to  achieve a
considerable decrease of the error of our numerical scheme in these
critical neighborhoods.

The evolution equations for $w_{_\Delta}$ and $h_{_\Delta}$ can be deduced
immediately by making use of the assumptions that $w=w_{_\Delta}+w_0$ and
$h=h_{_\Delta}+h_0$ satisfy (\ref{hyp}). The linearity of the involved
differential operators yields that 
\begin{equation}
{\partial_T}\Phi_{_\Delta}=A_R({\partial_R}\Phi_{_\Delta})+B_{_\Delta},
\label{hyp2}
\end{equation}
where 
\begin{equation}
B_{_\Delta}=B-{\partial_T}\Phi_0+A_R({\partial_R}\Phi_0)
\end{equation}
with $B$ being the functional as it is given by (\ref{coeff}) but 
evaluated at the functions $w=w_{_\Delta}+w_0$ and $h=h_{_\Delta}+h_0$,
moreover, $\Phi_{_\Delta}$ and $\Phi_0$  denote the vectors  
\begin{equation}
\Phi_{_\Delta}=\left(
\begin{array}{l}
{ w}_{_\Delta} \\
({ w}_{_\Delta})_T \\
({ w}_{_\Delta})_R \\
{ h}_{_\Delta}  \\
({ h}_{_\Delta})_T \\
({ w}_{_\Delta})_R \\
\end{array}\right), \ \ \ \ \ \ 
\Phi_0=\left(
\begin{array}{r}
{ w}_0 \\
{\partial_T w}_0 \\
{\partial_R w}_0 \\
{ h_0} \\
{\partial_T h}_0 \\
{\partial_R h}_0 \\
\end{array}
\right).
\end{equation} 
It is straightforward to verify that in the particular case when $w_0$
and $h_0$ are solutions of (\ref{2ymhe24}) and (\ref{2ymhe14}), 
i.e. $\Phi_0$ is a solution of (\ref{hyp}),  
\begin{equation}
B_{_\Delta}=B-B_0=\left(
\begin{array}{c}
{(w_{_\Delta}})_T \\
b_{w}-b_{w_0} \\
0 \\
({h_{_\Delta}})_T \\
b_{h}-b_{h_0}  \\
0 \\
\end{array}
\right),
\end{equation}
where $b_{w}$ and $b_{h}$ are supposed to be evaluated at the
functions ${w}={w_{_\Delta}+w_0 }$ and ${h}={h_{_\Delta}+h_0}$,
respectively. In all our numerical simulations, presented in this
paper,  $w_0$ and $h_0$ were chosen to be the static BPS solution
given by (\ref{ws}) and (\ref{hs}) with $C=g=1$.

\subsection{The treatment of the grid boundaries}\label{num}

In any numerical simulation the appropriate treatment of the grid
boundaries requires the most careful considerations. In the case
studied here the grid boundary consists of two disjoint parts, the
gridpoints representing the origin and the right edge lying in the
non-physical spacetime region.
It is the evaluation of various derivatives of the
dependent variables at these boundaries what desire very careful
investigations. The handling of these issues, as they are
implemented in our code, are described in the following subsections.

\subsubsection{The boundary at the origin}

\begin{figure}[ht]
\unitlength1cm
\centerline{
\begin{minipage}[t]{8.cm}
 \centerline{
  \epsfysize=4.5cm
  \epsfbox{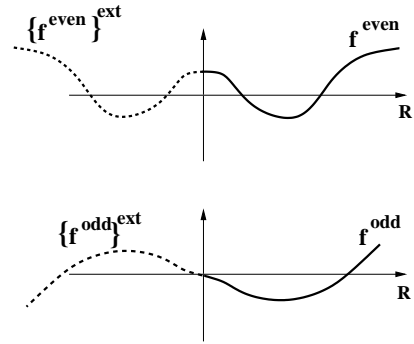}
 }
\caption{\footnotesize \label{orig} { 
The way of extension of functions to $R<0$ is indicated 
for even, $f^{{even}}$, and odd, $f^{{odd}}$, functions,
respectively. In fact, the extensions are determined by
either an axial or a central reflection to the $R=0$ line or to the
`central point' as indicated by the formulas $\{f^{even}\}^{ext}\vert_{-i}
=f^{even}_i$ and $\{f^{odd}\}^{ext}\vert_{-i}=-f^{odd}_i$, respectively.}}
\end{minipage}
}
\end{figure}

As it was already mentioned in Section\,\ref{smooth} at the
origin we can use spherical symmetry and the smoothness of the
field variables to gain information about the parity of the
relevant functions. This knowledge can then be used to extend these
functions onto a {\it virtually} enlarged grid. In particular,
for any function which can be divided into two parts, one of which has
a definite parity in coordinate $R$, while the other part is explicitly
known, suitable symmetry transformations can be used to define values
of the function at the additional grid points to the left of
the word line representing the origin (see Fig.\,\ref{orig}).
By making use of these values the required number of $R$-derivatives of the
applied variables can be determined  at $R=0$. More explicitly, as we
will see any function $f$ in our interest can be given as a sum of two
functions, $f=f^{_{d.p.}} + f^{_{e.k.}}$, one of 
which has a definite parity $f^{_{d.p.}}$, being either even or odd
function of the $R$-coordinate, and another explicitly known one
$f^{_{e.k.}}$. Such a function can
be extended, its extension will be denoted by $\{ f\}^{{ext}}$, to the
additional gridpoints as follows. Suppose, first, that  $f^{_{d.p.}}$
is an even function. Then $f^{_{even}}=f- f^{_{e.k.}}$ can be extended
to the left of $R=0$ simply by taking the axial reflection,
$\{f^{_{even}}\}^{{ext}}=(f-f^{_{e.k.}})^{AR}$, of its graph to the
$R=0$-axis, defined by the relation $(f^{_{even}})^{{AR}}\vert_{-i}=
f^{_{even}}_i$. Finally, by making use the fact that $f^{_{e.k.}}$
is explicitly known and thereby its extension is supposed to be also
known (potential singularities at the origin has already been
excluded by former assumptions), the extension
$\{f\}^{{ext}}$ of $f$ can be given as $\{ f\}^{{ext}}=
(f-f^{_{e.k.}})^{AR} + \{f^{_{e.k.}}\} ^{{ext}}$. A similar process
applies whenever $f^{_{d.p.}}$ is an odd function with the distinction
that axial reflection should be replaced by a `central reflection'
determined by the relation $(f^{_{odd}})^{{CR}} \vert_{-i}=
-f^{_{odd}}_i$. Thereby the extension of $f$ then can be given as
$\{f\}^{{ext}} = (f-f^{_{e.k.}})^{CR} + \{f^{_{e.k.}}\}^{{ext}}$.

Turning back to our concrete field variables note first that in the time
integration process we need to evaluate $R$-derivatives of 
the dependent variables listed as components of the vector valued
functions $\Phi$, see the relation (\ref{coeff}). Since $w$
and $h+rH\hspace{-.08cm}{}_{_\infty}$ were found to be even functions
of the original 
$r$-coordinate, it follows by a straightforward substitution that
\begin{eqnarray}
{ w}&=&{ w}^{^{even}},\\ 
{w}_T&=&{ w}_T^{^{even}},\\ 
{ w}_R&=&{w}_R^{^{odd}},\\ 
{ h}&=&{ h}^{^{even}}- \frac{R}{\Omega}H\hspace{-.08cm}{}_{_\infty},\\ 
{ h_T}&=&{ h}_T^{^{even}},\\ 
{h}_R&=&{h}_R^{^{odd}}- \frac{1+R^2}{2\Omega^2}H\hspace{-.08cm}{}_{_\infty}.
\end{eqnarray}
Thereby the extensions of these functions, to the enlarged grid,
yielded by the above described general process are
\begin{eqnarray}
{ w}^{\ \,l}_{-i}&=&{ w}^{l}_{i},\\
{w_T}^{\ \,l}_{-i}&=&{ w_T}^{l}_{i},\\ 
{ w_R}^{\ \,l}_{-i}&=&-{w_R}^{l}_{i},\\ 
{ h}^{\ \,l}_{-i}&=&{ h}^{l}_{i}+ 2H\hspace{-.08cm}{}_{_\infty}
\left(\frac{R}{\Omega}\right)^{l}_{i},\\    
{ h_T}^{\ \,l}_{-i}&=&{ h_T}^{l}_{i},\\ 
{h_R}^{\ \,l}_{-i}&=&-{h_R}^{l}_{i}-
H\hspace{-.08cm}{}_{_\infty}\left(\frac{1+R^2}{\Omega^2}\right)^{l}_{i}.
\end{eqnarray} 

Note that whenever we consider the evolution of $\Delta \Phi$ with
respect to a solution $\Phi_0$ of (\ref{hyp}) -- $\Phi_0$ possesses the
same parity properties as the hypothetical solution $\Phi$ itself --
thereby the components of $\Delta \Phi$ are necessarily even
functions of the $R$-coordinate so their extensions at the origin are
straightforward.

\subsubsection{The grid boundary in the non-physical spacetime}

The appropriate treatment of the boundary in the non-physical
spacetime requires 
completely different type of considerations. Clearly, at this part of
the boundary there is no way to enlarge the grid based on a suitable
combination of certain smoothness and symmetry requirements, as it was
possible to be done at the origin. Instead to be able to determine the
required number of $R$-derivatives we used the following two ideas:

First, at the edge and neighboring gridpoints we applied a numerical
adaptation what would be called as `one sided derivatives' (see
appendix B) in analytic investigations. By making use of these
approximations the time  integration process can be carried out along
the gridlines close to the edge. Second, our problem inherently is a
boundary initial value problem (see e.g. \cite{gko}). Thus appropriate
boundary conditions have to be chosen at the grid boundary to be able
to carry on the time integration scheme. There is a considerable
freedom in choosing boundary conditions at the edge of the grid. We
intended to choose the one which allows waves to travel from the left
toward the edge of the  grid without being reflected, moreover, which
excludes waves coming from beyond this edge toward the direction of
the domain of computation. The corresponding restrictions in terms of
our basic variables in the static hyperboloidal gauge are
\begin{eqnarray}
w_T&=&-\frac{1}{2}(R+1)^2 w_R,\\   
h_T&=&-\frac{1}{2}(R+1)^2 h_R. 
\end{eqnarray}
Numerical experiments justified that this choice did really ensure
that the above two requirements were satisfied.

\subsection{Specific choices of parameters in the code}

In order to concentrate sufficient number of grid points to
the central region where the monopole lives, as well as, to 
have enough  
grid points to resolve the expanding shell structures at large 
radius we have chosen $\kappa=0.05$ as the parameter included
in the coordinate transformation (\ref{TR21}) and (\ref{TR22}).
This specific choice turned out to be appropriate for all the 
simulations presented in this paper.

For convenience, the total number of spatial grid points was always
chosen to be an integer power of $2$. 
Our minimal resolution was $2^8=256$, with $4$ points in the negative
$R$ region and $16$ points in the $R>1$ domain.
To preserve the grid point positions when doubling the resolution,
in general, we used $4/256=1/64$ part of the total spatial grid points
for the negative $R$ mirror image points, and $16/256=1/16$ part of 
the total points for the unphysical domain.
Keeping the ``size'' of the unphysical region above $R>1$ to be
constant is useful when investigating the stability and the 
convergence of the code in that region, while the unnecessary points 
in the negative $R$ domain only decrease the speed of the code 
slightly.

In order to investigate the stability conditions of the numerical code 
it is helpful to write out the coordinate velocity of the radial null
geodesics.
Using (\ref{nullgeod}) we get
\begin{equation}
\frac{dR}{dT}=\pm\frac{1}{2}(1\pm R)^2 \ ,
\end{equation}
where the positive signs correspond to outgoing, the negative  signs
to ingoing geodesics.  The maximal coordinate velocity occurs at null
infinity $R=1$, taking  the value $2$ for outgoing geodesics.  The
absolute value of the coordinate velocity of ingoing geodesics  is
always smaller than $1/2$.  In principle this would allow any
numerical time step $\Delta T$  which is smaller than half of the
radial step $\Delta R$.  However, the evolution equations of the
massive field components  have an apparent singularity at $R=1$, where
our numerical evolution code was not appropriately stable unless the
time step was chosen as small as  $\Delta T=\Delta R/8$.

Choosing the parameter $\sigma$ in the dissipative term 
(\ref{dissip}) to be $\sigma=0.01$ stabilize our numerical code 
without influencing the results significantly even at lower 
resolutions.

\section{Testing the numerical code with massive and massless
  Klein-Gordon fields}\label{tests} 

In order to ascertain the appropriateness of our numerical code
we employed it to a physical system for which the time evolution
can be established by an independent, certainly more precise 
method.
Because of the linearity of the equations describing the massive 
Klein-Gordon field its time evolution can be calculated by the
Green function method.
If the Klein-Gordon field and its derivative are given on a spacelike
hypersurface then the field value can be calculated at any point 
in the future as a sum of two definite (numerical) integrals 
(for more details see, e.g., \cite{fr1}).
The study of the Klein-Gordon field is especially important,
since for several physically important systems, including the 
magnetic monopole, at large radius the various field components 
decouple, and behave like independent massive or massless 
linear Klein-Gordon fields. 

In the spherically symmetric case the Klein-Gordon field $\Phi$ 
satisfies the wave equation
\begin{equation}
\partial_r^2\Phi+\frac{2}{r}\partial_r\Phi-\partial_t^2\Phi
=m^2\Phi \ .
\end{equation}
Introducing the new field variable $z=r\Phi$, in the coordinate system 
$(T,R)$ defined by (\ref{TR21}) and (\ref{TR22}) the 
Klein-Gordon equation takes the form
\begin{equation}
\label{kgeq}
{\mathfrak P}z
= \frac{(R^2+1)^2}{4\Omega^2} m^2 z \ ,
\end{equation}
where the differential operator $\mathfrak P$ is defined
in (\ref{difop}).
Similarly to the monopole case the equation can be transformed to
a system of three first order partial differential equations by 
introducing the new dependent variables $z_T=\partial_T z$ and 
$z_R=\partial_R z$.

The initial data on the $T=0$ hypersurface was chosen as a 
specification of nonzero time 
derivative in a localized region superimposed on the vacuum value of
the field $z=0$ as
\begin{equation} \hskip -.27cm(z_T)_\circ = \left\{
\begin{array} {rr}  
    \frac{c}{\kappa} \exp\left[\frac{d}{(r-a)^2-b^2}\right], 
    & {\rm if}\ r\in
    [a-b,a+b] ;\\[4mm]
  0 \  , & {\rm otherwise},  \end{array} \right.\label{ffz}
\end{equation}
with the constants selected to be $a=2$, $b=1.5$, $c=70$, $d=10$
and $\kappa=0.05$ in this section.

\begin{figure}[ht]
\unitlength1cm
\centerline{
\begin{minipage}[t]{8.cm}
 \centerline{
  \epsfxsize=8.5cm 
  \epsfbox{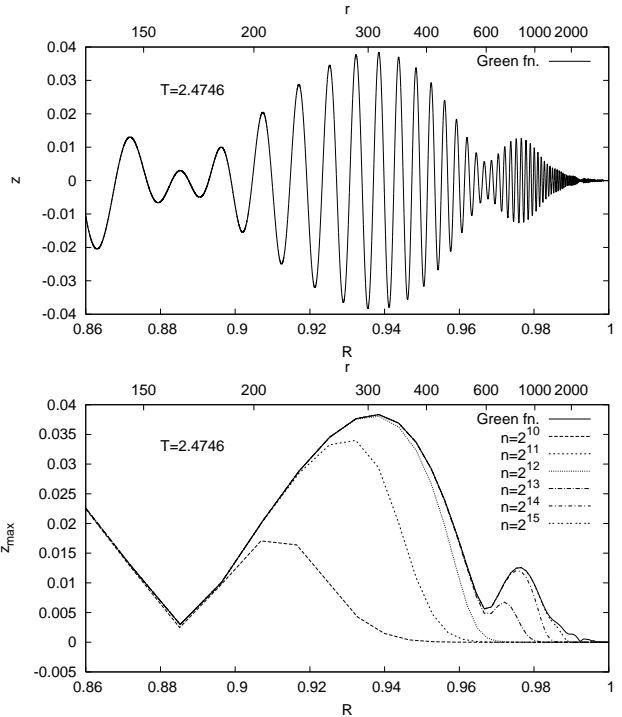}
 }
\caption{ \label{kg1}   
{The behavior of the massive Klein-Gordon field with $m=1$ at the
  $T=2.4746$ slice for large radiuses is shown.
In order to make the expanding shell structures more apparent the
function $z=r\Phi$ is plotted instead of $\Phi$. 
On the upper graph the oscillations of $z$ as calculated by the
Green function method is shown.
On the lower graph the upper envelope of the oscillations in $z$,
i.e. the curve  
connecting the maximum points of the function $z$, is shown for 
the Green function calculation, and also for the evolution code
corresponding to the indicated spatial resolutions. 
 }}  
\end{minipage}
}
\end{figure}

\begin{figure}[ht]
\unitlength1cm
\centerline{
\begin{minipage}[t]{8.cm}
 \centerline{
  \epsfxsize=8.5cm 
  \epsfbox{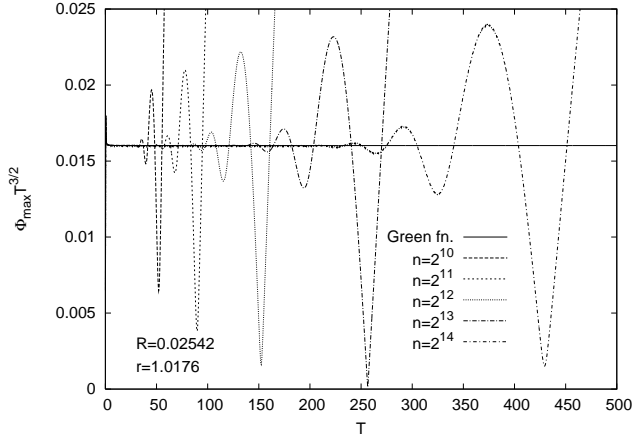}
 }
\caption{ \label{kg2}   {
The upper envelope curve of the oscillations of the Klein-Gordon field
$\Phi$ at a constant radius close to the center of symmetry. 
In order to compensate for the fast decay of the field, the envelope
of $\Phi T^{3/2}$ is plotted, which should tend to a constant
value on theoretical grounds.
Apart from a short initial period, the envelope of the Green 
function result is really constant.
The envelopes of the functions obtained by the evolution code
stay near this constant value for longer and longer times
as the number of spatial grid points is increasing.
 }}  
\end{minipage}
}
\end{figure}

\begin{figure}[ht]
\unitlength1cm
\centerline{
\begin{minipage}[t]{8.cm}
 \centerline{
  \epsfxsize=8.5cm 
  \epsfbox{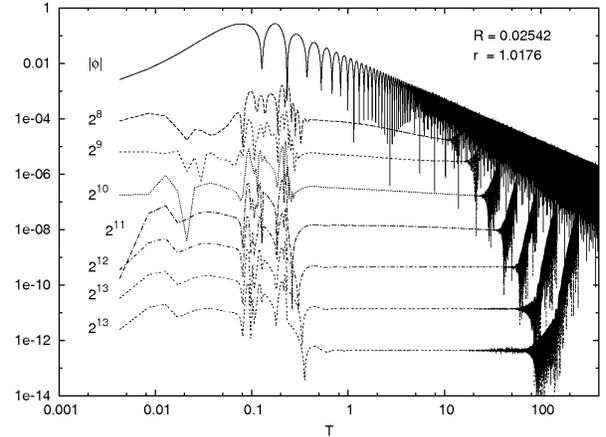}
 }
\caption{ \label{kg3}   {
Logarithmic plot of the absolute error of the value of the
massive Klein-Gordon field $\phi$ at constant radius $R=0.02542$ 
calculated with several different numerical resolutions 
corresponding to the indicated number of spatial grid points.
The correct absolute value of the field is also plotted in order 
to show the time intervals where the error is smaller than the
actual function value.
We note that the downward pointing peaks indicate moments of time
where the functions change signature.
In order to reduce the complexity of the figure, for each
resolution the error is plotted only up to a time where it becomes
close to the exact value.
 }}  
\end{minipage}
}
\end{figure}

First, we applied our evolution code to the massive 
Klein-Gordon field with $m=1$.
On Fig.\,\ref{kg1} we compare the field values on a constant $T$
slice obtained with different spatial resolutions to the precise
value calculated by the Green function method.
The chosen time slice is at $T=2.4746$, well after the null 
geodesic emanating from the origin at $T=0$ has reached null
infinity at $T=1$.  
It is apparent, that even if the results obtained by the evolution
code are not completely satisfactory close to $\scrip$ ($R=1$) the
field 
values in the central area remain correct for considerably longer
time intervals. 
On Fig.\,\ref{kg2} the upper envelope curve for $\Phi$ is shown
at a fixed radius, $R=0.02542$, corresponding to $r=1.0176$.
The frequency of the field oscillations approaches (from above) 
$m=1$ in terms of the original time coordinate $t$, which in
the $T$ coordinate corresponds to $20$. 
We can see, that as it may be expected from the behavior of the
ingoing null geodesics, the doubling of the resolution increases
the time interval where the numerical solution is valid 
approximately by a factor of two.
On Fig.\,\ref{kg3} the absolute value of the error at the same
constant $R$ radius is shown for several different numerical 
resolutions.
Initially, up to approximately $T=0.1$, the error values decrease
according to the expected fourth order convergence of the code.
Later, for $T>0.3$ the convergence become even faster,
approximately fifth order.
This behavior is due to dominance of the artificial dissipation
term which decreases as $(\Delta R)^5$.
At later stages a numerical instability arises and the error
increases until its absolute value reaches the magnitude
of $\phi$ and the numerical simulation is no longer reliable.

\begin{figure}[ht]
\unitlength1cm
\centerline{
\begin{minipage}[t]{8.cm}
 \centerline{
  \epsfxsize=8.5cm 
  \epsfbox{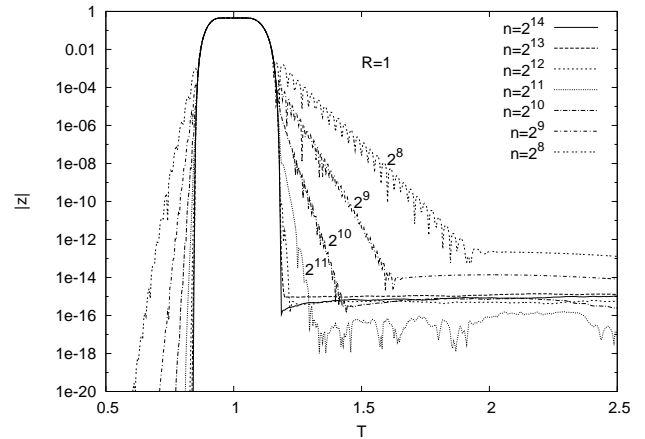}
 }
\caption{ \label{wsc1}   {
Time dependence of the absolute value of the rescaled field variable 
$z=r\Phi$ at null infinity $R=1$ obtained by our numerical code in
case of the massless Klein-Gordon field with 
numerical resolutions corresponding to the indicated number of spatial 
grid points.
 }}  
\end{minipage}
}
\end{figure}

We have also tested our code by applying it to the massless 
Klein-Gordon field with the same parameters in the initial data 
(\ref{ffz}). 
With $m=0$ the Klein-Gordon equation (\ref{kgeq}) is regular at 
null infinity $R=1$ in the compactified representation. 
On Fig.\,\ref{wsc1} we show the rescaled field variable $z=r\Phi$ 
at null infinity as a function of the hyperboloidal time coordinate
$T$ obtained with several different numerical resolutions.
Since in the massless case information spreads strictly with the
velocity of light, the value of $z$ has to be exactly zero
before the outgoing light ray from the outer edge of the nonzero
initial data, i.e. from $T=0$ and $r_b=a+b$, reaches null infinity, 
and has to return to zero again after the ingoing ray from the same 
point, bouncing through the center reaches null infinity as well.
This means that the signal can be nonzero only within the time
interval $\sqrt{\kappa^2r_b^2+1}-\kappa r_b < T 
< \sqrt{\kappa^2r_b^2+1}+\kappa r_b$ which, after 
substituting the values of the constants $\kappa$ and $r_b$,
corresponds to the interval $0.8402 < T < 1.1902$.
We see that the higher the resolution is the better the
numerical simulation can follow the abrupt changes in the magnitude
of the field variable.
Other feature that can be seen from the figure is that after the 
signal left the system a numerical noise remains, with a magnitude
of about $10^{-16}$ part of the size of the initial data.
The size of this noise remains in this low range for very long time 
intervals, even for $T\approx 10^4$. 

In order to keep the numerical noise at a low level for very long
time intervals a smooth cutoff in the artificial dissipation
term has been introduced in the unphysical region $R>1$.
This was necessary because of the incompatibility of the sixth 
derivative dissipative term with the outgoing boundary condition 
applied at the outer edge of the grid.

\section{Numerical results for the time evolution of magnetic
  monopoles}\label{numrez}

\subsection{Choice of initial data}

The numerical simulations presented in this paper were
carried out in the $\lambda=0$ case, when the Higgs
field is massless and the static monopole solution is
given by $w_s$ and $H_s$ in (\ref{ws}) and (\ref{hs}).
For our numerical simulations we used initial data which
consisted of a concentrated spherically symmetric pulse
superposed on the static monopole solution.
Defining $h_s=r(H_s-H_\infty)$, at $T=0$ we choose
$(w)_\circ=w_s$, $(w_R)_\circ=\partial_R w_s$,
$(h)_\circ=h_s$, $(h_R)_\circ=\partial_R h_s$,
$(h_T)_\circ=0$ 
and
\begin{equation} \hskip -.27cm(w_T)_\circ = \left\{
\begin{array} {rr}  
    \frac{c}{\kappa} \exp\left[\frac{d}{(r-a)^2-b^2}\right], 
    & {\rm if}\ r\in
    [a-b,a+b] ;\\[4mm]
  0 \  , & {\rm otherwise}.  \end{array} \right.\label{indat}
\end{equation}
In all of the following simulations we take $a=2$, $b=1.5$
and $d=10$, but we choose different amplitudes $c$.
As it is indicated by Table.\,\ref{inidata}, for instance,
for our basic choice $c=70$, which is the case discussed
in \cite{fr2}, the energy provided by the pulse
is about $55\%$ of the energy of the background static 
monopole solution. 
\begin{table} [h]
	  \centering
\begin{tabular}{|c||c|c|}
\hline
$c$ & $E_c$ & $E_c/E_s$ \\
\hline
\hline
0.7 & 0.000696595 & 0.00005543 \\
\hline
7 & 0.0696595 & 0.005543 \\
\hline
70 & 6.96595 & 0.5543 \\
\hline
280 & 111.455 & 8.869 \\
\hline
1120 & 1783.28 & 141.9 \\
\hline
\end{tabular}
\caption{ \label{inidata}   { The value of the energy $E_c$ of the
exciting pulse, along with the value of its ratio $E_c/E_s$ with
respect to the energy, $E_s=12.56637$, of the original static BPS
solution is given
for several values of the parameter $c$. In fact, the listed values of $c$ are
labeling exactly those excitations for which the properties of the time
evolution are going to be discussed below.   
}}
\end{table}
Notice, however, that for $c=280$ the energy of the pulse gets to be
$16$-times of the energy of $E_{c=70}$ which implies that the energy
of the pulse is $887\%$ of the energy of the original static
BPS monopole. Clearly such an excitation is far too strong to be
considered as a  perturbation and, as we shall see below, highly
non-linear aspects of the evolution show up in this case.

\subsection{The qualitative picture}

Before presenting an accurate and comprehensive quantitative analysis
of the dynamics of the excited monopole it is informative to consider
the qualitative aspects of the investigated processes. This is done in
this section by making use of {\it spacetime plots} of various
quantities, where the term `spacetime plot' is to indicate that the
quantity in question is given as a function of space and time always
over a rectangular coordinate domain of the $(R,T)$
plane. Accordingly, on these spacetime plots the horizontal lines will
always represent the $T=const$  coordinate lines, along which the
value of the $R$-coordinate is changing between the indicated minimal
and maximal values, $R_{min} \leq R \leq R_{max}$, increasing from the
left to the right. Similarly, the vertical lines are the $R=const$
coordinate  lines, along which the value of the $T$-coordinate is
changing between the indicated initial and final values, $T_i \leq T
\leq T_f$, increasing from the bottom to the top.

\subsubsection{The evolution of the basic variables}

This short subsection is to present some spacetime plots informing
about the  behavior of the basic variables $w$ and $h$ as the
monopole react for weak and strong excitations.
\begin{figure}[ht]
\unitlength1cm
\centerline{
\begin{minipage}[t]{8.cm}
 \centerline{
  \includegraphics[width=8cm]{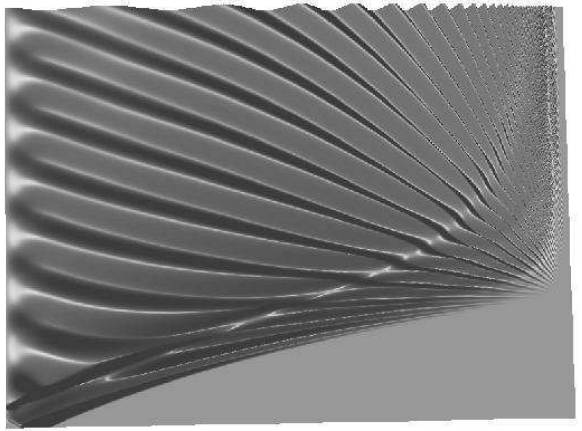}
 }
\caption{\footnotesize \label{wd07}   { The spacetime plot of the
    difference $w-w_0$ between the dynamical and the static values of
    the field variable $w$  is given for the weak excitation with
    $c=0.7$ while $T$ and $R$ take values from the intervals $0<T<
    2.9661$ and $0\leq R\leq 1$. The maximum, the minimum and the
    average values of $w-w_0$ are $0.004230$, $-0.002398$ and
    $9.021\cdot 10^{-6}$. }}
\end{minipage}
}
\end{figure}
\begin{figure}[ht]
\unitlength1cm
\centerline{
\begin{minipage}[t]{8.cm}
 \centerline{
  \includegraphics[width=8cm]{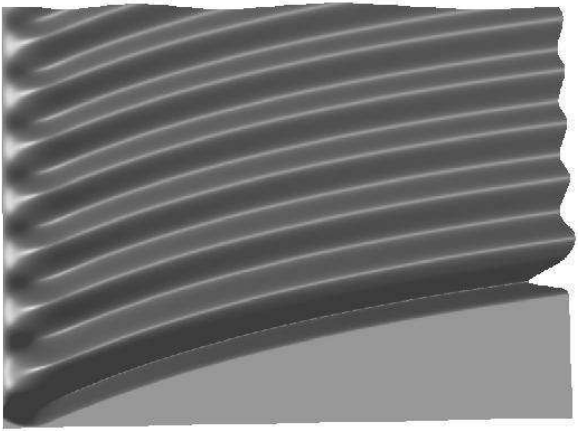}
 }
\caption{\footnotesize \label{hd07}   { The spacetime plot of the
difference $h-h_0$ between the dynamical and the static values of the
field variable  $h$ is given for the weak excitation with $c=0.7$
while $T$ and $R$ take values from the intervals $0\leq T\leq  2.9661
$ and $0\leq R\leq 1$. The maximum, the minimum and the average values
of $h-h_0$ are $0.0009632$,  $-0.004673$ and $-0.0001738$. }}
\end{minipage}
}
\end{figure}
\begin{figure}[ht]
\unitlength1cm
\centerline{
\begin{minipage}[t]{8.cm}
 \centerline{
  \includegraphics[width=7.9cm]{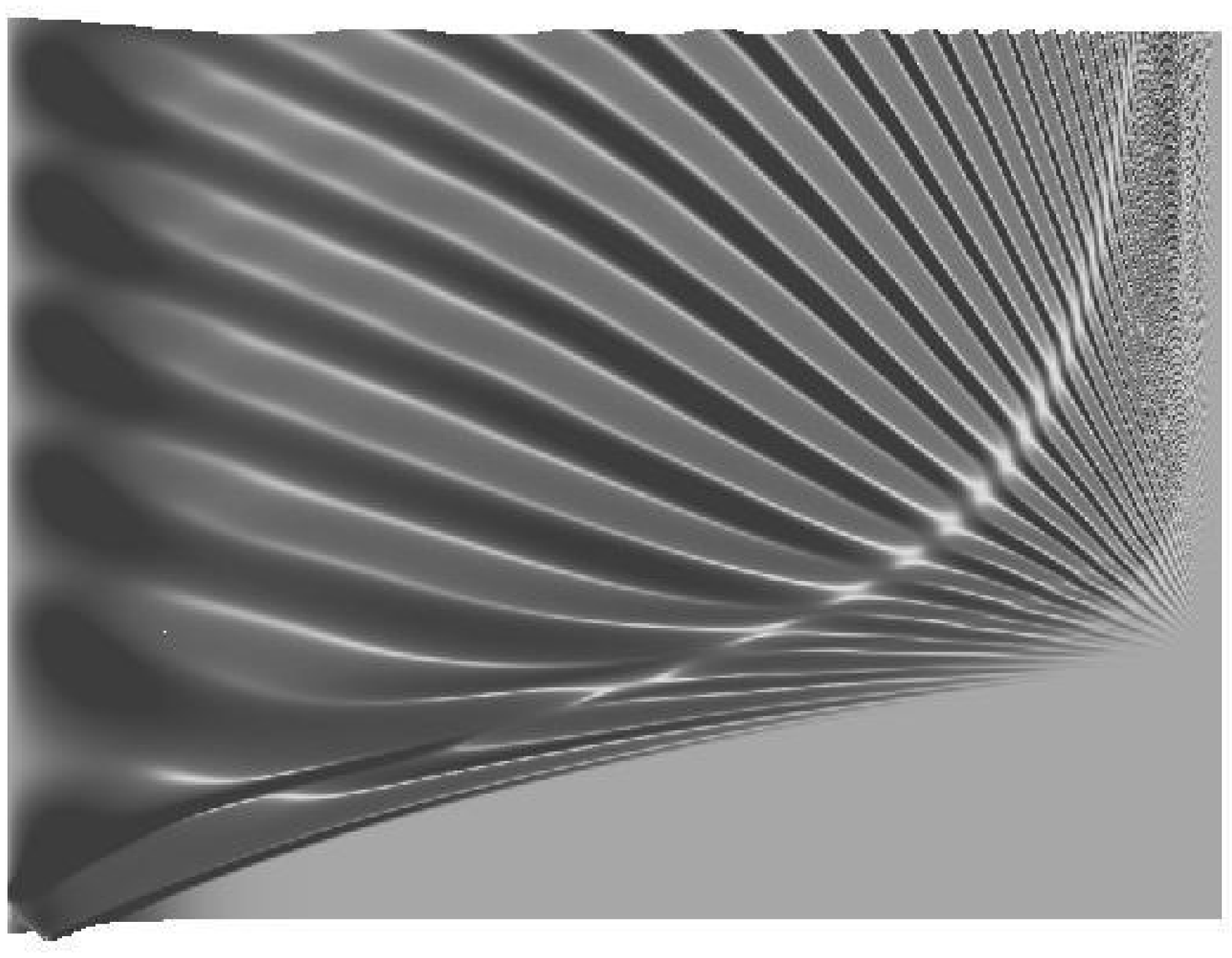}
 }
\caption{\footnotesize \label{w280}   { The spacetime plot of $w$
is given for the strong excitation with $c=280$ while $T$ and $R$ take
values from the intervals $0\leq  T\leq  2.9661 $ and $0\leq R\leq
1$. The maximum, the minimum and the average values of $w$ are
$2.070$,  $-1.210$ and $0.07346$.  }}
\end{minipage}
}
\end{figure}
\begin{figure}[ht]
\unitlength1cm
\centerline{
\begin{minipage}[t]{8.cm}
 \centerline{
  \includegraphics[width=7.9cm]{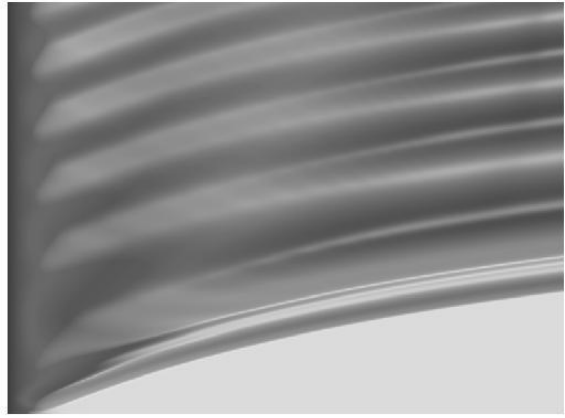}
 }
\caption{\footnotesize \label{h280G}   { The spacetime plot of $h$
is given for the strong excitation with $c=280$ while $T$ and $R$ take
values from the intervals $0\leq  T\leq  2.9661 $ and $0\leq R\leq
1$. The maximum, the minimum and the average values of $h$
are $-5.317$, $0$ ,$-3.034$. }}
\end{minipage}
}
\end{figure}

There are immediate similarities and differences to be observed. First of
all, the basic features indicated by Figs.\,\ref{wd07} and \ref{hd07}
and Figs.\,\ref{w280} and \ref{h280G} are similar although the energy
of the exciting pulse relative to the energy of the initially static
BPS monopole is negligible for the weak excitation -- that is why not
the field values $w$ and $h$ themselves but their difference with
respect to the static background are plotted on Figs.\,\ref{wd07} and
\ref{hd07} -- while it is about nine times of the energy of the BPS
monopole for the strong excitation. In spite of the immediate
similarities the frequencies of the developing oscillations at the
center are different. It is a bit of surprise that the larger the
excitation is the lower is the initial frequency. Comparing the $w$ and $h$
evolutions, the most
important differences get manifested if one considers the shape
of the  curved lines associated with the maximum and minimum field
values. While these curves seem to approximate ingoing null
geodesics in case of the massive Yang-Mills field, see
Figs.\,\ref{wd07} and \ref{w280}, they closely follow outgoing null
rays in case of the massless Higgs field as indicated by
Figs.\,\ref{hd07} and \ref{h280G}. Since all physically
interesting quantities like the energy, energy current or magnetic
charge densities are derived from $w$ and $h$, along with their
derivatives, these derived quantities shall show a mixture of these
clear characters as it will be seen below.

These figures make it also transparent that the two basic
variables $w$ and $h$, although, are strongly coupled in the
central region, they can be considered as
being independent far away from there.  

\subsubsection{The spacetime dependence of energy and energy
  transfer}\label{enentr} 

One might expect that the most natural quantity to be applied when one
tries to present the evolution of a system is
the spacetime plot of the energy density. Such a quantity has a meaning
only after the family of observers measuring it has been specified. In
our setting it is natural to consider the energy density  measured by
static observers represented by the unit timelike Killing vector field
$t^a=(\partial/\partial t)^a$.  Let us denote by $\varepsilon$ the associated 
energy density which can be given by the relation
$\varepsilon=T_{ab}t^a t^b$.

\begin{figure}[ht]
\unitlength1cm
\centerline{
\begin{minipage}[t]{8.cm}
 \centerline{
  \includegraphics[width=8.7cm]{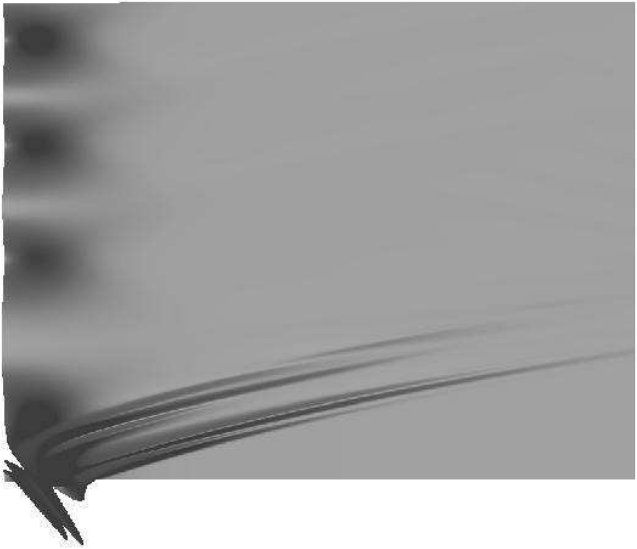}
 }
\caption{\footnotesize \label{epsilon}   { The spacetime plot of the
    energy density $\varepsilon=T_{ab}t^a t^b$ is shown
    for a strong excitation with amplitude $c=280$ and for the space and
    time intervals $0\leq R\leq 0.4915$ and $0\leq  T\leq 2.119$. 
    The maximum, the minimum and the average values of
    $\varepsilon$ are $13.33$, $2.214\cdot 10^{-6}$ and
    $0.03845$.   }}
\end{minipage}
}
\end{figure}

Fig.\,\ref{epsilon} is included to make it clear that
even for a large excitation the information concerning the dynamics of
the system is not properly reflected by the energy density. 
There is a clearly visible huge pulse at the beginning, the top of the
peak is pointing far out from the plane of the figure at the lower left
corner, which is followed by a moderated oscillation at the very
center. Although the range of the radial coordinate is now only
$0\leq R\leq 0.4915$, corresponding to $0\leq r\leq 25.92$, it is easy to agree
that beside the short dynamical starting nothing interesting is visible on most
of this figure.

Investigating the system it turned out that instead of monitoring the
behavior of the energy density $\varepsilon$ it is more
informative to consider the time evolution of the energy density
${\mathcal{E}}={\mathcal{E}}(T,R)$ `associated with shells of radius
$R$', defined by
\begin{equation}\label{calE}
{\mathcal{E}}=4\pi r^2\frac{dr}{dR}{T^0}_0=
2\pi\kappa(R^2+1)\frac{R^2}{\Omega^{4}}{T^0}_0\ ,
\end{equation}
where the components of the energy momentum tensor (\ref{Tab}) are calculated 
in the coordinate system $x^a=(T,R,\theta,\phi)$.
Justification of this definition will be given in subsection\,\ref{enba}, where
the energy conservation will be considered in details.
The integral, $\int^{1}_{0}{\mathcal{E}}dR$ with respect to the
$R$-coordinate gives, for any fixed value of $T$, the total energy
$E(T)$ of the $T=const$ hypersurface.

\begin{figure}[ht]
\unitlength1cm
\centerline{
\begin{minipage}[t]{8.cm}
 \centerline{
  \includegraphics[width=7.9cm]{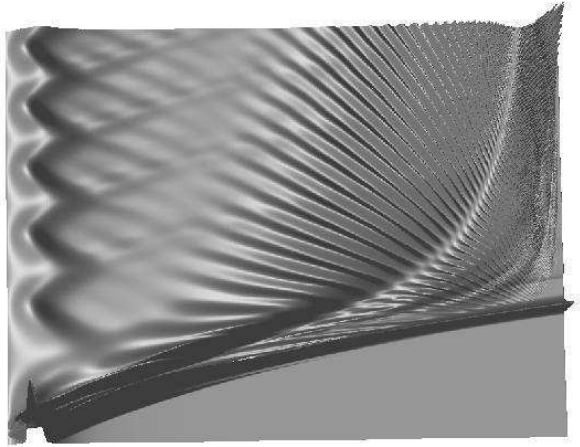}
 }
\caption{\footnotesize \label{t00s280}   { The spacetime plot of
    ${\mathcal{E}}$ for the large excitation with $c=280$ is shown for the
    space and time intervals $0\leq  R\leq 1$ and $0\leq T\leq 2.9661$. The
    maximum, the minimum and the average values of ${\mathcal{E}}$ are
    $5572$, $0$, $96.68$.   }}
\end{minipage}
}
\end{figure}

The advantages related to the use of ${\mathcal{E}}$, instead of
$\varepsilon$, are visible on Fig.\,\ref{t00s280}. This spacetime
plot of ${\mathcal{E}}$ is already very informative, it clearly
manifests the most important characteristic features of the beginning
of the evolution.
First there is a direct energy transport to $\scri^+$ (represented by
the right vertical edge) by the massless Higgs field following
outgoing null geodesics. The second phenomenon is that right after a
short period a quasi-stable oscillation starts at the central region
(close to the left vertical edge). Finally, the formation of the
expanding shells of high  frequency oscillations in the distant region
is also clearly visible. The time dependence of the mean value, the
amplitude and  frequency of some of the oscillating quantities will be
investigated in details in subsection\,\ref{freq}.

To understand the evolution of the investigated system it is
also important to consider spacetime plots giving information about the energy
transfers represented by the
energy current density `associated with shells of radius $R$',
\begin{equation}\label{calS}
{\mathcal{S}}=4\pi r^2\frac{dr}{dR}{T^1}_0=
2\pi\kappa(R^2+1)\frac{R^2}{\Omega^{4}}{T^1}_0\  .
\end{equation}
Referring to the results contained by subsection\,\ref{enba}, it can also be
justified that ${\mathcal{S}}$ is the quantity, the time integral
$\int^{T_1}_{T_2}{\mathcal{S}}dT$ of which, for any fixed value
of $R$, gives the energy  passing though the $R=const$ hypersurface
during the time interval $T_1\leq T \leq T_2$.

\begin{figure}[ht]
\unitlength1cm
\centerline{
\begin{minipage}[t]{8.cm}
\centerline{
  \includegraphics[width=8.1cm,angle=0]{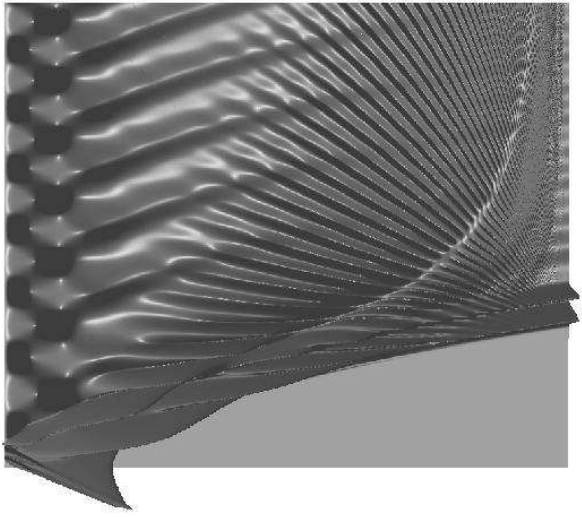}
 }
\caption{\footnotesize \label{t10s1}   { The spacetime plot of
the quantity ${\mathcal{S}}$, for the intermediate 
    excitation with amplitude $c=70$ is shown for the space and time
    intervals $0\leq R \leq 1$ and $0\leq T\leq 2.88559$. 
}} 
\end{minipage}
}
\end{figure}

Fig.\,\ref{t10s1} shows the spacetime plot of ${\mathcal{S}}$ for the
case of the intermediate energy exciting pulse with $c=70$. The same
characteristics of the evolution as on the previous figure, relevant for the
large exciting pulse, show up clearly on this figure, as well.


In virtue of Fig.\,\ref{t10s1} at least two different phenomena
clearly manifest themselves. Firstly,
the amplitude of the energy current density is much smaller beyond the
shells of high frequency oscillations (this region is represented by
the relatively smooth part close  to the right vertical edge) than
before reaching these shells. Actually, the `ribs', representing the
maximal values of ${\mathcal{S}}$ which  apparently follow the shape
of outgoing null rays all the way  out to $\scrip$ but they are
getting to be modulated more and more by the slowly moving  massive
shells of oscillations in the distance. Secondly, it is also apparent
that some  sort of inward directed energy transfer starts at the
location where the `ribs' meet the massive shells of oscillations. 

\begin{figure}[ht]
\unitlength1cm
\centerline{
\begin{minipage}[t]{8.cm}
\centerline{
  \includegraphics[width=8cm,angle=0]{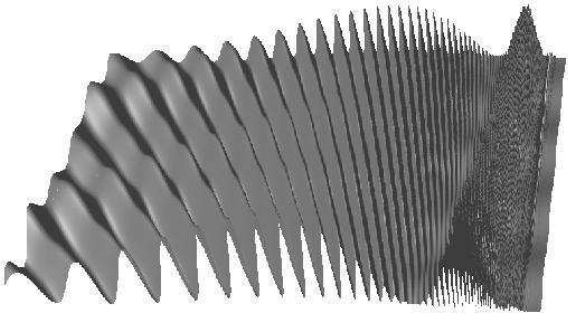}
 }
\caption{\footnotesize \label{t10s2}   { The
spacetime plot of the quantity ${\mathcal{S}}$, for the intermediate
excitation with $c=70$ is shown for the space and time intervals
$0.5829 \leq R \leq 1$ and $2.11864\leq T\leq 2.88559 $.  }}
\end{minipage}
}
\end{figure}

These shells of oscillations can be nicely pictured by focusing to a
small subsection of the previous spacetime diagram as shown on
Fig.\,\ref{t10s2}. 

\begin{figure}[ht]
\unitlength1cm
\centerline{
\begin{minipage}[t]{8.cm}
\centerline{
  \includegraphics[width=7cm,angle=0]{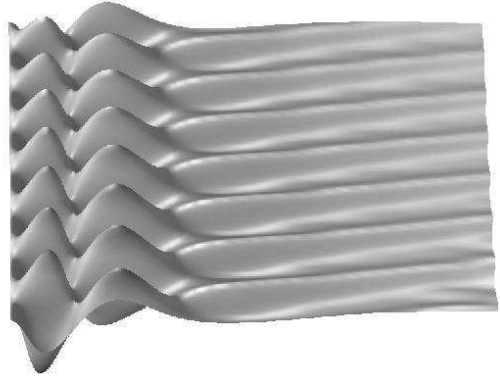}
 }
\caption{\footnotesize \label{t00sd70Gk}   { The space and time
    dependence of the ${\mathcal{E}}-{\mathcal{E}}_0$ for the
    intermediate  excitation with amplitude $c=70$ is shown for the
    coordinate domain given by the relations $0 \leq R \leq 0.35$ and
    $ 1.695 <T< 4.237$.  }}
\end{minipage}
}
\end{figure}

\begin{figure}[ht]
\unitlength1cm
\centerline{
\begin{minipage}[t]{8.cm}
\centerline{
  \includegraphics[width=7cm,angle=0]{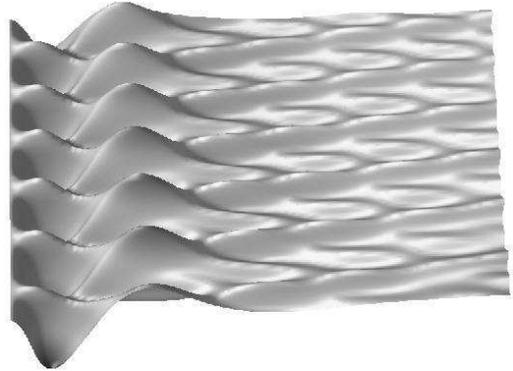}
 }
\caption{\footnotesize \label{t00sd280Gk}   { The spacetime plot of
    ${\mathcal{E}}-{\mathcal{E}}_0$, for the large 
    excitation with amplitude $c=280$ is shown for the space and time
    intervals $0 \leq R \leq 0.35$ and $1.695 \leq T\leq 4.237$. The maximum
    values are about ten times larger than the those are on
    Fig.\,\ref{t00sd70Gk}. 
}}
\end{minipage}
}
\end{figure}

\begin{figure}[ht]
\unitlength1cm
\centerline{
\begin{minipage}[t]{8.cm}
\centerline{
  \includegraphics[width=7cm,angle=0]{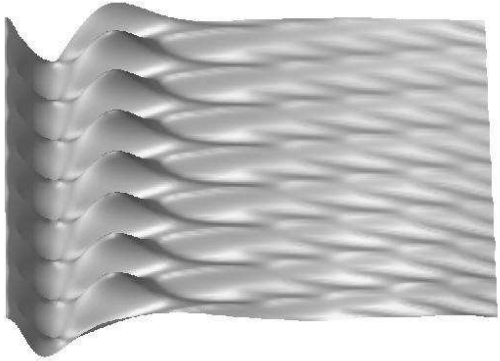}
 }
\caption{\footnotesize \label{t00sd1120Gk}   { The spacetime plot of
    ${\mathcal{E}}-{\mathcal{E}}_0$, for the extra large  excitation
    with amplitude $c=1120$ is shown for $0 \leq R \leq 0.35$ and for
    a later time interval $36.441 \leq T\leq 38.983$. Here the
    maximum values are still more than three times 
    larger than those on Fig.\,\ref{t00sd70Gk}. 
}}
\end{minipage}
}
\end{figure}

Finally, we would like to attract attention to the appearance of some
interesting features of the 
oscillations close to the origin. As the strength of the applied
excitation is increased, the associated shapes, which can be clearly
recognized  on the following spacetime plots, are getting more and
more complex. In virtue of
Figs.\,\ref{t00sd70Gk}, \ref{t00sd280Gk} and
\ref{t00sd1120Gk}, it is tempting to say that there might be a mean of
switching on more and  more oscillating degrees of freedom of the
monopole as the energy of the exciting pulse is increased.
Notice that despite the relatively late time interval the oscillating shapes are
still present.
Since ${\mathcal{E}}={\mathcal{E}}_0=0$ at the left vertical edge corresponding
to $R=0$, it can be clearly seen from the figures that for the larger
excitations the energy contained in the central region is definitely smaller
than the energy contained in the same region of the static monopole.
This means that at the centre still there is a huge deficit in the energy,
i.e. considerably large fraction of the energy of the static BPS
monopole is still somewhere in the outer region.
A very large pulse in this sense almost destroys the monopole, scattering its
energy to faraway regions. Of course, because of the charge conservation
this energy has to come back slowly to the center, forming the well localized
monopole again with smaller and smaller amplitude oscillations.

\subsubsection{The space and time dependence of the pressure}

There are two types of pressure, the radial and tangential, that can be
associated with our spherically symmetric system. They are given as
$P_{rad}=T^{ab}n^{(R)}_a n^{(R)}_b$ and $P_{ang}=T^{ab}n^{(\theta)}_a
n^{(\theta)}_b$, respectively, where $n^{(R)}_a$ and $n^{(\theta)}_a$
denote the unit norm spacelike vector proportional to the coordinate
differentials $(dR)_a$  and $(d\theta)_a$, respectively. The radial
and tangential pressures are not equal to each other as it may happen
in case of generic spherically symmetric configurations.  Note that
both of these pressures are identically zero for the static BPS
configuration. 

\begin{figure}[ht]
\unitlength1cm
\centerline{
\begin{minipage}[t]{8.cm}
\centerline{
  \includegraphics[width=8cm,angle=0]{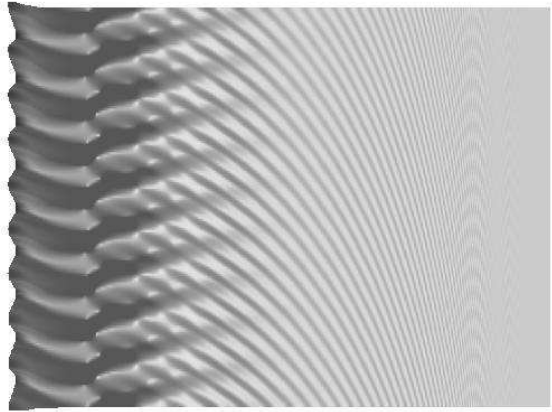}
 }
\caption{\footnotesize \label{prG}   { The time evolution of the
radial pressure, $P_{rad}$, is shown for the intermediate exciting
pulse with amplitude  $c=70$. The coordinate domain is given by the
relations $0 \leq R \leq 1$ and $16.949 \leq T\leq 19.915 $.  Notice
that this time period refers to that part of the evolution when the
system behaves already much like a quasi-breather.      }}
\end{minipage}
}
\end{figure}

\begin{figure}[ht]
\unitlength1cm
\centerline{
\begin{minipage}[t]{8.cm}
\centerline{
  \includegraphics[width=8cm,angle=0]{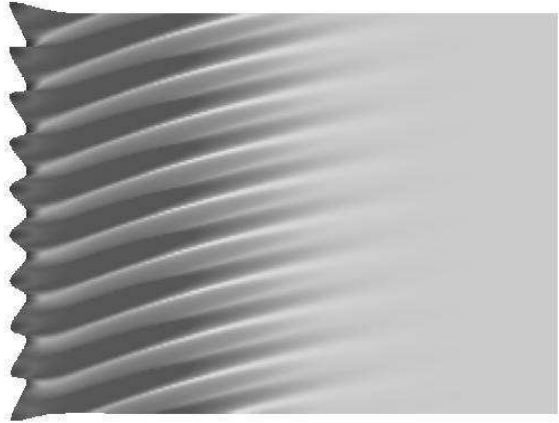}
 }
\caption{\footnotesize \label{paG}   { The time evolution of the
angular pressure, $P_{ang}$, is shown. The system and the coordinate
domain is exactly the same as on Fig.\,\ref{prG}.
}}
\end{minipage}
}
\end{figure}

To be able to show the tiny features far form the center on these
spacetime plots, on Figs.\,\ref{prG} and \ref{paG}, we needed to multiply the
pressures with an enormous scaling factor, and use an almost orthogonal
projection (i.e.\ very faraway point of view) to compensate the scaling factor.
Actually, the ratios of the values of the radial and angular pressure at the
origin and at the middle of the
plot, at $R=0.5$, are about  $1.5\cdot 10^4$  and $2.3\cdot 10^4$,
respectively. This means that both of the pressures are significantly
larger at the origin, where the monopole lives, than anywhere else.  

Although there is a noticeable phase difference between the almost
sinusoidal oscillations of the two pressures at the origin, moreover, clear 
differences also show up in the shape of the oscillations further away, no
really surprisingly new phenomenon, in addition to the formerly reported ones,
can be observed on these figures.

\subsection{The quantitative picture}

Before presenting a more detailed description of the behavior of the
fundamental physical quantities let us show some simple but convincing
results concerning the numerical preservation of the constraint
equations during the investigated long term evolutions.          

\subsubsection{Monitoring of the constraints}

\begin{figure}[ht]
\unitlength1cm
\centerline{
\begin{minipage}[t]{8.cm}
 \centerline{
  \epsfxsize=8.5cm 
  \epsfbox{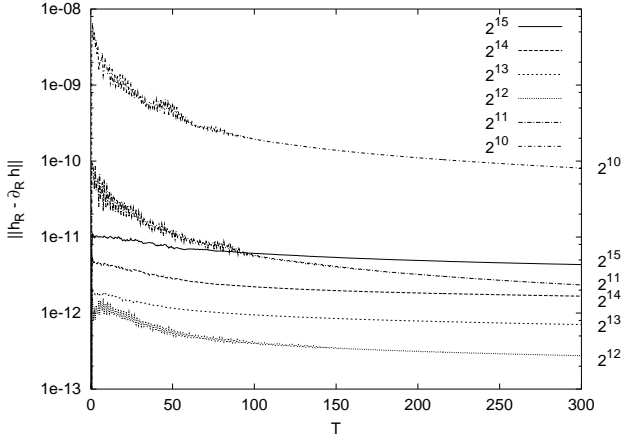}
 }
\caption{ \label{ch}   {
Time dependence of the $h$ constraint violation for 
several numerical resolutions corresponding to the
indicated number of spatial grid points. The amplitude
parameter in the initial data was chosen to be $c=70$.
 }}  
\end{minipage}
}
\end{figure}

Recall that the $R$-derivatives of $w$ and $h$ are used as dependent
variables, in consequence of which, 
the relations $w_R=\partial_R w$ and $h_R=\partial_R h$ play the role
of constraints now.  In the analytic setting these relations are
preserved by the evolution equations, provided that they hold on the
initial surface.  One of the possible tests of our numerical code is the
monitoring of the  violation of these constraints.  On Fig.\,\ref{ch}
the $L^2$ norm of the $h$ constraint is presented as a function of
time, defined as
\begin{equation}
||h_R-\partial_R h||=\left(\int_{0}^{1}\left(
h_R-\partial_R h\right)^2\,dR\right)^{\frac{1}{2}}\ . 
\end{equation}
It can be seen from the figure that for lower resolutions the error
decreases when one increases the number of grid points.  However, at
higher resolutions, most likely because of rounding errors, the
constraint violation becomes higher again.  It is also apparent that
apart from a short initial period the constraint violation is a
decreasing  function of time.  The time dependence of the $w$
constraint violation is similar to that of $h$ on Fig.\,\ref{ch}, only
its magnitude is somewhat smaller.

Finally, we would like to emphasize that the numerical preservation of
the constraint equations, reported above, in our rather long time
evolutions is considerably good. Our code seem to be free, at least in
case of the investigated system, from the deficiencies of other
approaches, where in case of various dynamical systems exponential
increase of the numerical violation of 
the constraint equations have been observed.

\subsubsection{Energy balances }\label{enba}

\begin{figure}[ht]
\unitlength1cm
\centerline{
\begin{minipage}[t]{8.cm}
 \centerline{
  \epsfysize=6.5cm 
  \epsfbox{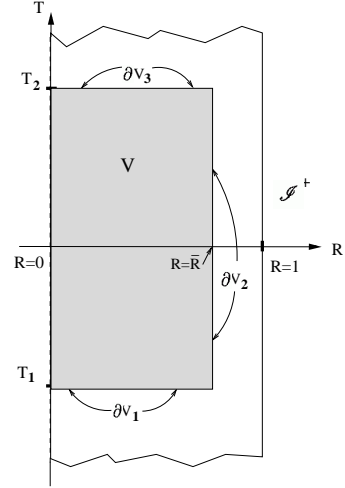}
 }
\caption{\footnotesize \label{eb}   {The boundary $\partial V=
\partial V_1 \cup \partial V_2 \cup \partial V_3$ of the shaded
spacetime domain, represented by the union of the $T=T_1$, $T=T_2$ and
$R=\bar R$ hypersurfaces is shown. }}  
\end{minipage}
}
\end{figure}

To monitor the appropriateness of the applied numerical scheme we
calculated energy balances for various spacetime domains. In
particular, we considered the type of domains shown by Fig.\,\ref{eb} 
bounded by $T=const.$ and $R=const.$ hypersurfaces.

To see how the energy balances can be calculated recall that whenever
$t^a$ is a Killing vector field, satisfying the Killing equation
$\nabla^{(a}t^{b)}=0$, the vector field 
\begin{equation}
j^a={T^a}_bt^b
\end{equation}
is divergence free, i.e. 
\begin{equation}
\nabla_aj^a=0 \,.
\end{equation}
Then, by making use of Stokes' theorem we have 
\begin{equation}
\int_V\nabla_aj^a\epsilon=\int_{\partial V}n_aj^a\tilde
\epsilon =0\,, \label{stokes}
\end{equation}
where $\epsilon$ denotes the 4-volume element while $\tilde \epsilon$
is the 3-volume element induced on the boundary $\partial V$ of
$V$. Note that $\tilde \epsilon_{abc}=n_e{\epsilon^e}_{abc}$ where $n_e$
is the (outward pointing) unit normal 1-form field on $\partial V$. 
For the following considerations we choose $t^a$ as the unit normed static
timelike Killing vector field $t^a=(\partial/\partial t)^a$.

Let us denote by $\mathcal{B}(\bar T,\bar R)$ the ball of radius $R=\bar
R$ centered at the origin of the $T=\bar T(=const)$ hypersurface,
moreover, by $\mathcal{C}(T_2,T_1,\bar R)$ the portion of the cylindrical
hypersurface $R=\bar R$ between the $T=T_1$ and $T=T_2$
hypersurfaces. With these notations the boundaries, as indicated on
Fig.\,\ref{eb},  $\partial V_1$ and $\partial V_3$ are the balls
$\mathcal{B}(T_1,\bar R)$ and $\mathcal{B}(T_2,\bar R)$, while
$\partial V_2$ is the cylinder $\mathcal{C}(T_2,T_1,\bar R)$
connecting them. Then the energy contained in a ball
$\mathcal{B}(T,R)$ can be given as 
\begin{equation}\label{sint}
E(T,R)=\int_{\mathcal{B}(T,R)} n_a^{(t)}j^a\tilde\epsilon^{(t)} \,,
\end{equation}
where $\tilde\epsilon^{(t)}$ is the volume element on the constant
$T$ hypersurface and $n_a^{(t)}$ is its future pointing normal
vector.
Similarly, the energy transported through the timelike boundary 
$\mathcal{C}(T_2,T_1, R)$ is given as 
\begin{equation}\label{tint}
S(T_2,T_1,R)=\int_{\mathcal{C}(T_2,T_1, R)} n_a^{(s)}j^a
\tilde\epsilon^{(s)} \,,
\end{equation}
where $\tilde\epsilon^{(s)}$ is the volume element on the constant
$R$ hypersurface and $n_a^{(s)}$ is its outwards pointing normal
vector.
Using these notations the energy balance equation (\ref{stokes})
takes the form
\begin{equation}\label{stokes2}
E(T_2,R)-E(T_1,R)+S(T_2,T_1,R)=0 \ .
\end{equation}

Note that while in analytic considerations the l.h.s. of
(\ref{stokes2}) is always identically zero, for any choice of $T_2,T_1$
and $R$, in meaningful numerical simulations the corresponding
quantity is merely close to zero, i.e. an apparent violation of the
energy conservation happens. In fact, the numerical value
$N_{vec}(T_2,T_1,R)$ of the
``violation of the energy conservation'' associated with a spacetime
domain $V$, which is defined as 
\begin{equation}\label{Nstokes2}
N_{vec}= E(T_2,R)-E(T_1,R)+S(T_2,T_1,R),
\end{equation}
can be used as one of the possible monitorings of the appropriateness
of a numerical code.  

To evaluate the integrals (\ref{sint}) and (\ref{tint}) we need to
determine the volume elements $\tilde\epsilon^{(t)}$ and
$\tilde\epsilon^{(s)}$ which can be given as special cases of the
relation 
\begin{equation}
\tilde\epsilon_{abc}=\sqrt{|g_{(4)}|}\varepsilon_{eabc}n^e
\end{equation}
where $\varepsilon_{eabc}$ denotes the Levi-Civita alternating tensor
with $\varepsilon_{0123}=1$. To get $\tilde\epsilon^{(t)}$ and
$\tilde\epsilon^{(s)}$ consider now the Minkowski
spacetime in coordinates $(T,R,\theta,\phi)$ associated with the
applied conformal representation. Then the components
$g_{\alpha\beta}$ of the metric tensor, $g_{ab}$, and its inverse read
as
\begin{equation}
g_{\alpha\beta}= \frac{1}{\Omega^2}\left(
\begin{array}{cccc}
\frac{\Omega^2}{\kappa^2} & R & 0 & 0  \\
R & -1 & 0 & 0 \\
0 & 0 & -R^2 & 0 \\
0 & 0 & 0 & -R^2 \sin^2\theta \\
\end{array}
\right) ,
\end{equation}
and
\begin{equation}
g^{\alpha\beta}= \Omega^2\left(
\begin{array}{cccc}
\frac{4}{(R^2+1)^2} & \frac{4R}{(R^2+1)^2} & 0 & 0  \\
\frac{4R}{(R^2+1)^2} & -\frac{4\Omega^2}{(R^2+1)^2\kappa^2}&
0  & 0\\ 
0 & 0 & -\frac{1}{R^2} & 0 \\
0 & 0 & 0 & -\frac{1}{R^2 \sin^2\theta} \\
\end{array}
\right).
\end{equation}
Calculating the determinant $g_{(4)}$ of the spacetime metric we
obtain
\begin{equation}
\sqrt{|g_{(4)}|}=\frac12\Omega^{-4} (R^2+1)R^2 \sin\theta.
\end{equation}

The future pointing normal form of $\mathcal{B}(T,R)$ has the components
$n_\alpha^{(t)}=(1/\sqrt{g^{00}},0,0,0)$.
This implies then that
$n^{(t)\,0}=g^{00}n_0^{(t)}=\sqrt{g^{00}}=\frac{2\Omega}{R^2+1}$,
moreover, we
have that
\begin{equation}\label{3vol}
\tilde \epsilon_{\alpha\beta\gamma}^{(t)}
= \frac{R^2\sin\theta}{\Omega^3}
\varepsilon_{0\alpha\beta\gamma}=\sqrt{|h^{(t)}|}(dR)_\alpha\wedge
(d\theta)_\beta\wedge (d\phi)_\gamma,
\end{equation}
where $h^{(t)}$ is the determinant of the three metric
$h_{ab}^{(t)}=g_{ab}-n_a^{(t)}n_b^{(t)}$ induced by $g_{ab}$ on
$\mathcal{B}(T,R)$,  i.e.
\begin{equation}
\sqrt{|h^{(t)}|}=\frac{R^2\sin\theta}{\Omega^3}
\end{equation}
on $\mathcal{B}(T,R)$.
Similarly, the normal of $\mathcal{C}(T_2,T_1,R)$, pointing out from
the domain $V$, is
$n_\alpha^{(s)}=(0,1/\sqrt{-g^{11}},0,0)$ and thereby
$n^{(s)\,1}=g^{11}n_1^{(s)}=-\sqrt{-g^{11}}
=-\frac{2\Omega^2}{\kappa(R^2+1)}$ which
implies, in particular, that 
\begin{equation}
\tilde \epsilon_{\alpha\beta\gamma}^{(s)}\hspace{-.04cm} =\hspace{-.04cm}
-\frac{R^2\sin\theta}{\kappa \Omega^2} 
\varepsilon_{1\alpha\beta\gamma}=\sqrt{|h^{(s)}|}(dT)_\alpha\wedge
(d\theta)_\beta\wedge (d\phi)_\gamma,
\end{equation}
where $h^{(s)}$ is the determinant of the induced metric
$h_{ab}^{(s)}=g_{ab}+n_a^{(s)}n_b^{(s)}$, and 
\begin{equation}
\sqrt{|h^{(s)}|}=\frac{R^2\sin\theta}{\kappa\Omega^2}.
\end{equation}

Taking the above relations into account we get that 
\begin{equation}
E(T,R) =\int_0^{ R} {\mathcal{E}} (T,R')\, {\rm d}R'\ ,
\end{equation}  
\begin{equation}
S(T_2,T_1,R)=\int_{T_1}^{T_2} {\mathcal{S}} (T,R) {\rm d}T\ ,
\end{equation}
where ${\mathcal{E}}$ and ${\mathcal{S}}$ are the functions defined in
(\ref{calE}) and (\ref{calS}).

\begin{figure}[ht]
\unitlength1cm
\centerline{
\begin{minipage}[t]{8.cm}
 \centerline{
  \epsfxsize=8.5cm 
  \epsfbox{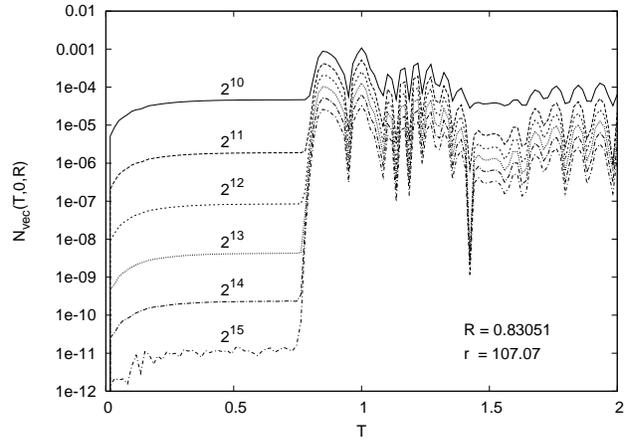}
 }
\caption{ \label{ec1}   {
Initial behavior of the numerical violation of the energy conservation  
$N_{vec}(T,0,R)$ for different spatial resolutions is shown.
The spacetime domain is bounded by the constant time surfaces $T_1=0$
and $T_2=T$, while the cylindrical boundary is at $R=0.83051$,
corresponding to $r=107.07$.
 }}  
\end{minipage}
}
\end{figure}

\begin{figure}[ht]
\unitlength1cm
\centerline{
\begin{minipage}[t]{8.cm}
 \centerline{
  \epsfxsize=8.5cm 
  \epsfbox{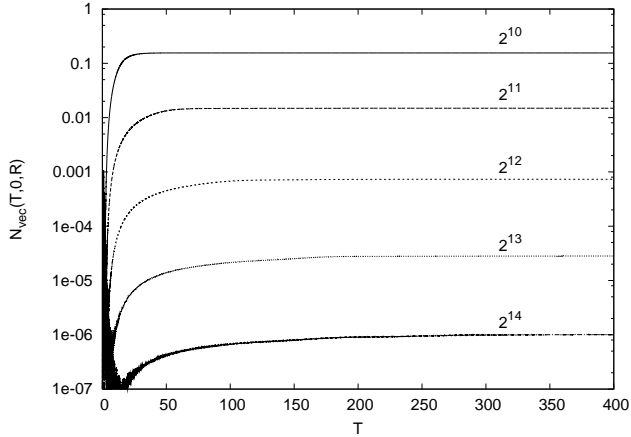}
 }
\caption{ \label{ec2}   { The time dependence of the violation of the
energy conservation $N_{vec}(T,0,R=0.83051)$  for the same numerical
simulations as on the previous figure is shown  but for a much longer
time interval.  }}
\end{minipage}
}
\end{figure}

On Figs.\,\ref{ec1} and \ref{ec2} the time dependence of the
violation of the energy conservation $N_{vec}(0,T,R)$ is shown for
numerical runs with various resolutions evolving from initial
data with amplitude $c=70$.
It can be seen from these figures that the error in the
energy conservation decreases according to expected fourth
order convergence during an initial period up to
approximately $T=0.7$ and also later, after $T>20$.
In the intermediate region the conservation violation
decreases more slowly.
This is possibly due to the highly oscillating character 
of the fields in that time interval, especially since the
energy current integrals were calculated only with a second
order convergent method.

\begin{figure}[ht]
\unitlength1cm
\centerline{
\begin{minipage}[t]{8.cm}
 \centerline{
  \epsfxsize=8.5cm 
  \epsfbox{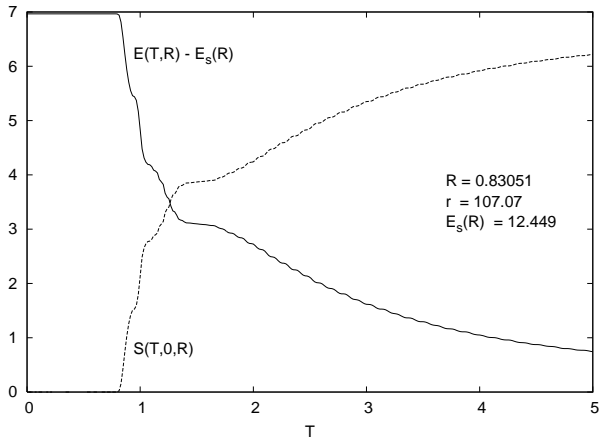}
 }
\caption{ \label{eisi}   {
The solid curve on the graph shows the time dependence of the
energy contained inside a radius $R$ on the constant $T$
hypersurface $E(T,R)$ decreased by the energy content of the
static monopole, $E_s(R)$.
The dashed curve represents the total energy transported through 
the constant $R$ hypersurface until time $T$.
The sum of these two curves is constant up to errors represented
by $N_{vec}(T,0,R)$. 
 }}  
\end{minipage}
}
\end{figure}

\begin{figure}[ht]
\unitlength1cm
\centerline{
\begin{minipage}[t]{8.cm}
 \centerline{
  \epsfxsize=8.5cm 
  \epsfbox{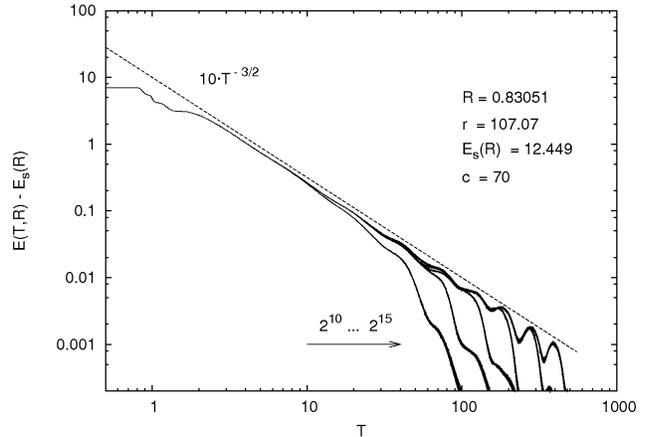}
 }
\caption{ \label{eid}   {
Time dependence of the extra energy content $E(T,R)-E_s(R)$
for the same system as on the previous figure, but now for
longer time periods and with logarithmic axes.
 }}  
\end{minipage}
}
\end{figure}

The magnitude of the presented constraint violations is really
meaningful only if it can be compared to the full energy
content.
On Fig.\,\ref{eisi} the time dependence of the energy 
content $E(T,R)$ and energy loss $S(T,0,R)$ is shown for
the same $R$ as on Figs.\,\ref{ec1} and \ref{ec2}.
To bring the two curves of the graph into the the same range,
instead of the total energy $E(T,R)$ the energy difference between
$E(T,R)$ and 
that of the static monopole solution $E_s(R)$ is shown on
Figs.\,\ref{eisi} and \ref{eid}.
For the chosen radius $R=0.83051$ the energy of the static background monopole
is $E_s(R)=12.449$, less than double of the energy provided by the initial 
deformation.

The time dependence of the difference $E(T,R)-E_s(R)$ of the dynamical
and static energy functions, $E(T,R)$ and $E_s(R)$, for longer time
periods is shown logarithmically 
on Fig.\,\ref{eid} for increasing numerical resolutions.
The convergence of the presented curves provides us a strong
indication on how long we can take our numerical results seriously.
For the highest resolution used, i.e. that with $2^{15}$ spatial
grid points, the calculation becomes unreliable after $T=300$.
We note that in physical time, i.e. in time measured in mass unites,
the time interval $\Delta T=300$, at the centre, corresponds
to $\Delta t=6000$,  meanwhile the central monopole performs nearly
one thousand oscillations.

\subsubsection{Linear and non-linear effects}\label{et} 

\begin{figure}[ht]
\unitlength1cm
\centerline{
\begin{minipage}[t]{8.cm}
 \centerline{
  \epsfxsize=8.5cm 
  \epsfbox{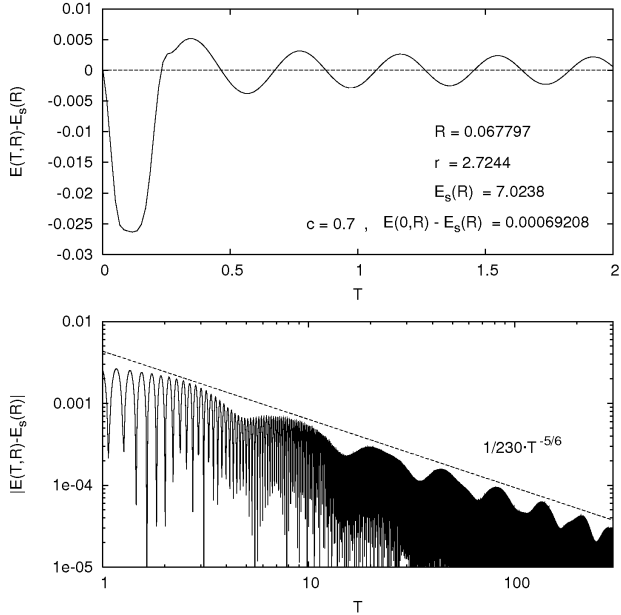}
 }
\caption{ \label{eid20a}   { The extra energy $E(T,R)-E_s(R)$ as a
function of time contained inside radius $R=0.067797$ is shown for an
initial and for a longer time period, $1\leq T \leq 300$. By a careful analysis
it can be justified that on average the energy content close to the origin might
be slightly lower than that of the static monopole. The forming of a
long lasting breathing phase with a clear asymptotic 
time dependence can also be clearly recognized.  }}
\end{minipage}
}
\end{figure}

\begin{figure}[ht]
\unitlength1cm
\centerline{
\begin{minipage}[t]{8.cm}
 \centerline{
  \epsfxsize=8.5cm 
  \epsfbox{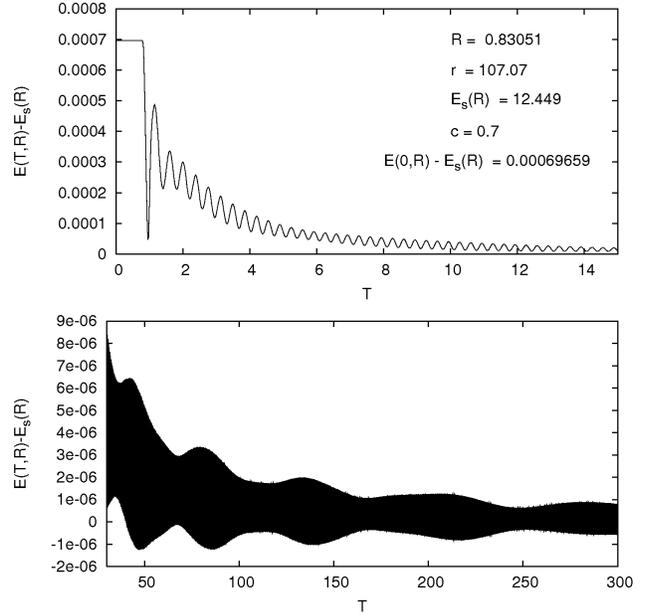}
 }
\caption{ \label{eid200a}   {
The extra energy content $E(T,R)-E_s(R)$ for radius $R=0.83051$
is presented 
in case of the low energy deformation, with $c=0.7$.  It is
transparent that, although the energy content close to the origin
might be slightly lower than that of the static monopole, up to this
relatively large radius the system possesses more energy on the 
average than the static monopole had.
 }}  
\end{minipage}
}
\end{figure}

In order to distinguish linear and nonlinear effects in the 
evolution of magnetic monopoles we first present results
corresponding to very small initial deformation, with amplitude
$c=0.7$. 
Then the extra energy provided by the initial pulse is
only $0.000696595$, which is very small compared to the energy 
of the static monopole, which is $12.56637$ to seven digits
precision.
On Fig.\,\ref{eid20a} the time dependence of the extra energy
is presented in a ball of radius $R=0.067797$, which
corresponds to $r=2.7244$.
Inside this radius the static background solution contains
energy $E_s(R)=7.0238$, which is about 56 percent of the
total energy of the static monopole.
For this low energy case, after a short initial period, the 
energy content starts oscillating around the energy of
the static monopole.
The time decay of these oscillations is $T^{-5/6}$ to a good
approximation.
On Fig.\ \ref{eid200a} the energy content is presented in 
a much larger ball, with radius $R=0.83051$, corresponding
to $r=107.07$.
Inside this large radius the energy contained in the static
monopole is $12.449$, which is $99$ percent of the total
static energy.
It is apparent that, because the Yang-Mills field is massive, part of
the energy provided by the pulse remains 
inside this radius for quite a long time.
The oscillations of the energy content will be centered
around the static value only after about $T>100$. 

\begin{figure}[ht]
\unitlength1cm
\centerline{
\begin{minipage}[t]{8.cm}
 \centerline{
  \epsfxsize=8.5cm 
  \epsfbox{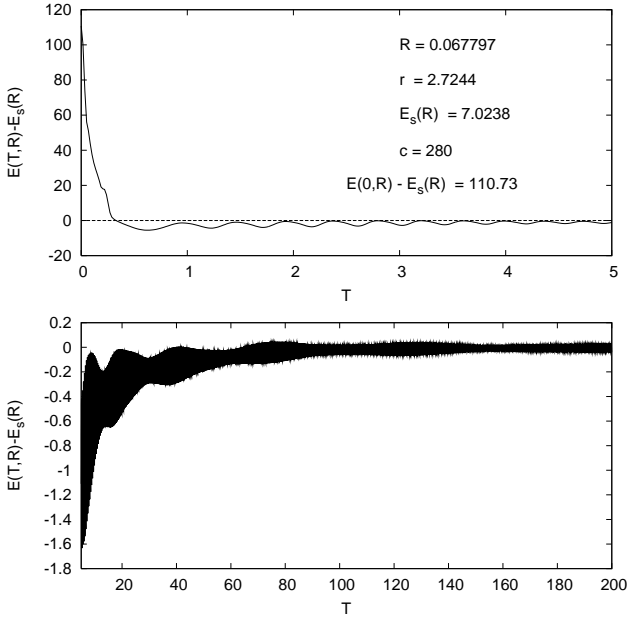}
 }
\caption{ \label{eid20b}   {
The extra energy content $E(T,R)-E_s(R)$ for radius $R=0.067797$
is shown for high energy deformation corresponding to $c=280$.  The graphs
  clearly indicate that for a large exciting pulse the average energy
  content of the central region is below that of the static monopole. 
 }}  
\end{minipage}
}
\end{figure}

\begin{figure}[ht]
\unitlength1cm
\centerline{
\begin{minipage}[t]{8.cm}
 \centerline{
  \epsfxsize=8.5cm 
  \epsfbox{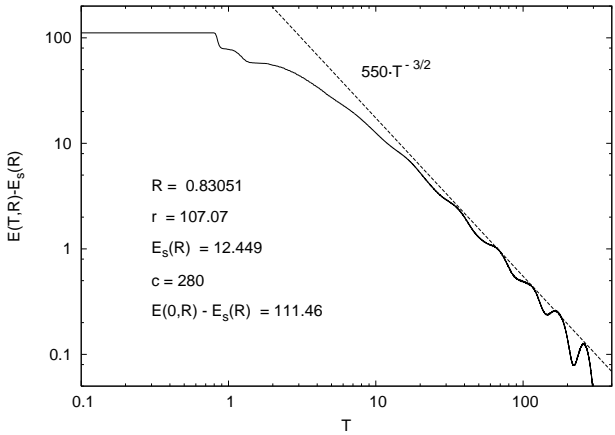}
 }
\caption{ \label{eid200b}   {The extra energy content $E(T,R)-E_s(R)$
is shown for the larger radius, $R=0.83051$, and for the high energy
deformation with $c=280$.  The asymptotic power law time decay can
easily be read off the {\it log scale} plot.  }}
\end{minipage}
}
\end{figure}

On Fig.\,\ref{eid20b} the energy contained in the same small
radius $R=0.067797$ as the one used at Fig.\,\ref{eid20a} is
shown for a large initial deformation corresponding to $c=280$.
In this case the energy provided by the pulse inside this radius is $110.73$,
which is about $15$ times more than the energy of the static monopole in this
region. The main difference compared to the previous case is
that the oscillation of the energy is not centered on the  static
monopole energy anymore.  The average value of the energy is below the
static value for a very long time period.  We interpret this as a
nonlinear effect.  The expanding initial pulse sweeps out not only the
provided extra energy but also a part of the energy of the static
monopole.  It takes a long time to get back this energy which is
stored in the massive Yang-Mills field oscillations at intermediate
distances from the monopole.  On Fig.\,\ref{eid200b} the extra energy
is shown for the same large initial pulse but for the larger radius,
$r=107.07$.
Inside this large radius, just like in case of the tiny
excitation with $c=0.7$ (see Fig.\,\ref{eid20a}), the energy remains
above the static value during the entire evolution where our numerical
simulation can be considered to be valid.

\begin{figure}[ht]
\unitlength1cm
\centerline{
\begin{minipage}[t]{8.cm}
 \centerline{
  \epsfxsize=8.5cm 
  \epsfbox{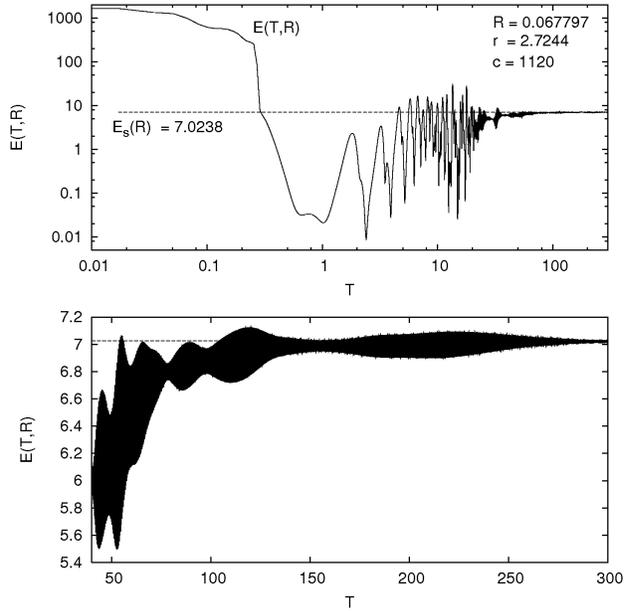}
 }
\caption{ \label{eid20n}   {
The full energy content $E(T,R)$ inside radius $R=0.067797$ is
indicated for an { enormously} large initial pulse with
$c=1120$. Even though the monopole seems to be swept out from the
central region for the initial period eventually the behavior of the
system returns to an oscillating normal monopole state.
 }}  
\end{minipage}
}
\end{figure}

On Fig.\,\ref{eid20n} the effect of an extra large initial
pulse is presented.
In this case the amplitude of the initial pulse was chosen to be $c=1120$,
corresponding to provided energy $1771.75$ inside the $R\leq0.067797$ region.
In this case the initial pulse quickly sweeps out most of
the energy of the monopole from the central region.
However, after some highly energetic oscillations, the
major part of the monopole energy returns, and the energy 
content starts to oscillate around a value below that of the
static monopole, { like in the case of} the less energetic
nonlinear excitations considered previously. 

\subsubsection{The time dependence of the total radiated energy at
  $\scrip$}\label{et1} 

\begin{figure}[ht]
\unitlength1cm
\centerline{
\begin{minipage}[t]{8.cm}
 \centerline{
  \epsfxsize=8.5cm 
  \epsfbox{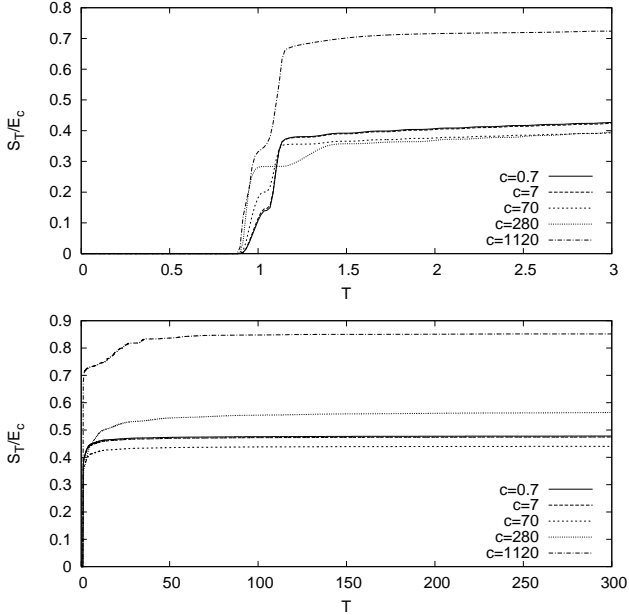}
 }
\caption{ \label{sintSP1} { The ratio $S_{T}/E_c$ is plotted for five
different choices of the initial amplitude $c$, where $S_{T}=S(T,0,1)$
is the energy radiated to infinity up to $T$ and $E_c$ denotes the
energy provided by the exciting pulse.  }}
\end{minipage}
}
\end{figure}

\begin{figure}[ht]
\unitlength1cm
\centerline{
\begin{minipage}[t]{8.cm}
 \centerline{
  \epsfxsize=8.5cm 
  \epsfbox{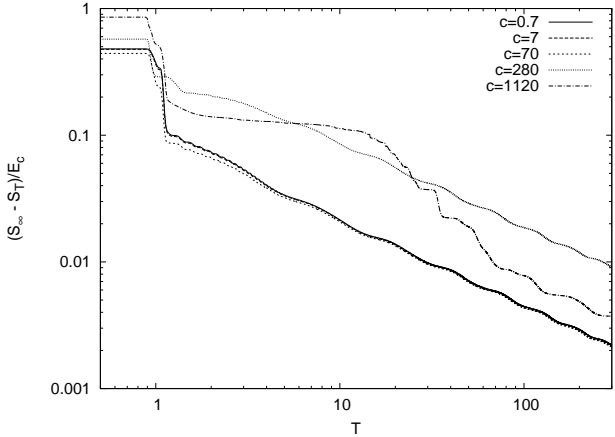}
 }
\caption{ \label{sintSP2} { The quantity $(S_\infty-S_{T})/E_c$ is
plotted logarithmically for the indicated values of the initial
amplitude $c$. }}
\end{minipage}
}
\end{figure}

\begin{figure}[ht]
\unitlength1cm
\centerline{
\begin{minipage}[t]{8.cm}
 \centerline{
  \epsfxsize=8.5cm 
  \epsfbox{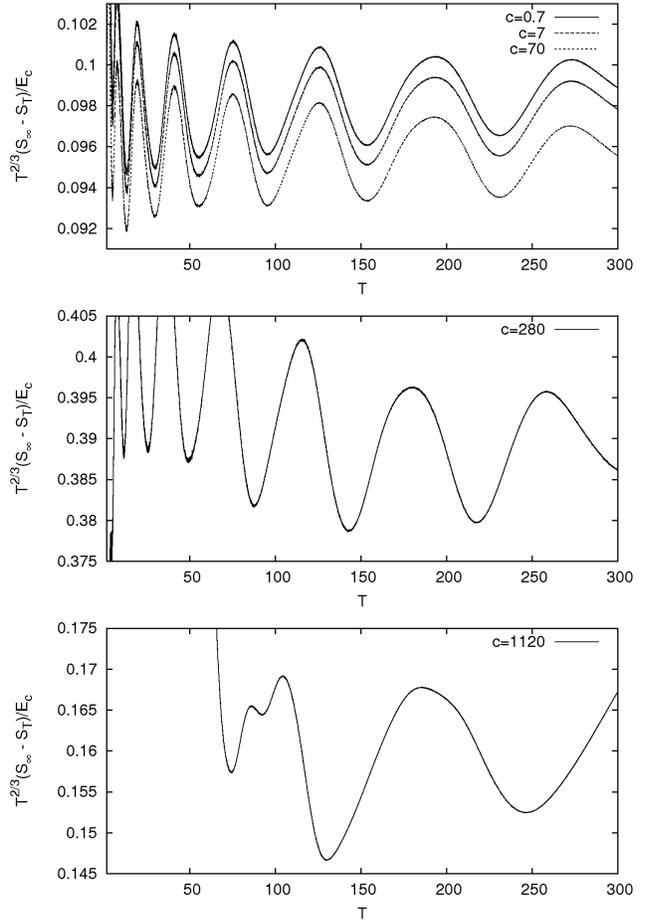}
 }
\caption{ \label{sintSP3} { The quantity $(S_\infty-S_{T})/E_c$
multiplied by $T^{2/3}$ is plotted for the five different initial data
configurations.}}
\end{minipage}
}
\end{figure}

In this subsection we provide a short account on the time dependence
of the energy radiated to future null infinity.  Since the Yang-Mills
field is massive, $w$ decays exponentially when approaching the $R=1$
line corresponding to null infinity. This implies that the intensity
of the radiation to $\scrip$ is determined only by $h$ and its first
derivatives at $R=1$.  The full radiated energy up to the moment $T$,
i.e. $S_{T}=S(T,0,1)$ in terms of the notation introduced in section
\ref{enba}, depends on the amount of energy $E_c$ provided by the
initial deformation (\ref{indat}). For our special choice of initial
data $E_c$ is proportional to $c^2$.  On Fig.\ \ref{sintSP1} the ratio
$S_{T}/E_c$ is plotted for five different initial data amplitudes.
It can be seen that except for very large initial energies the
radiated energy is closely proportional to the energy contained in the
initial pulse. In Table \ref{sintt}, for the five chosen $c$ amplitude
value, we list the energy of the initial pulse $E_c$, its ratio to the
energy of the static monopole, $E_c/E_s$, the fraction of the initial
energy radiated up to $T=300$, and the fraction of the initial energy
radiated during the whole time evolution.
\begin{table} [ht]
	  \centering
\begin{tabular}{|c||c|c|c|c|}
\hline
$c$ & $E_c$ & $E_c/E_s$ & $S_{300}/E_c$ & $S_\infty/E_c$ \\
\hline
\hline
0.7 & 0.000696595 & 0.00005543 & 0.47793 & 0.48014\\
\hline
7 & 0.0696595 & 0.005543 & 0.47428 & 0.47646\\
\hline
70 & 6.96595 & 0.5543 & 0.44037 & 0.44250\\
\hline
280 & 111.455 & 8.869 & 0.56388 & 0.5725\\
\hline
1120 & 1783.28 & 141.9 & 0.85177 & 0.8555\\
\hline
\end{tabular}
\caption{ \label{sintt}{ The energy $E_c$ of the initial pulse, its
ratio to the static monopole energy, i.e.\ $E_c/E_s$, furthermore, the
fraction of the energy radiated to $\scrip$ up to time $T=300$ and
$T=\infty$, are given for five different values of the initial
amplitude parameter $c$. }}
\end{table}
The value $S_\infty$ was estimated by assuming that the energy content
of the system decreases according to a power law decay with $T^{-3/2}$
during the later stages of the radiation process. As it can be seen on
Figs.\ \ref{sintSP2} and \ref{sintSP3}, this is certainly a very good
approximation for weaker initial excitations, and it appears to be
consistent with the late time behavior of more energetic evolutions.

Notice that since $S_\infty<E_c$ the limit of the energy associated
with the system, on $T=const$ hypersurfaces, {\it does not} tend to
the energy of the original BPS monopole. This observation is in fact
in accordance with the existence of the massive shells of high
frequency oscillations. These shells never reach future null infinity,
and hence they store for the rest of the entire evolution a part of
the energy of the original exciting pulse. This phenomenon gets to be
even more transparent if one thinks of a simple massive Klein-Gordon
system starting, say, from a trivial configuration where only the
exciting pulse stores energy. The pulse disperses, as it is expected,
but the system is conservative so the total energy has to be
conserved. Due to this, there is no upper bound on the increase of the
frequency associated with these shells of oscillations forming even in
this simplest possible linear case. Let us mention that a more
detailed discussion of this phenomenon can be found in \cite{fr1}.

\subsubsection{The space and time dependence of the energy transfer
  towards $\scrip$}\label{et2}

\begin{figure}[ht]
\unitlength1cm
\centerline{
\begin{minipage}[t]{8.cm}
 \centerline{
  \epsfxsize=8.5cm 
  \epsfbox{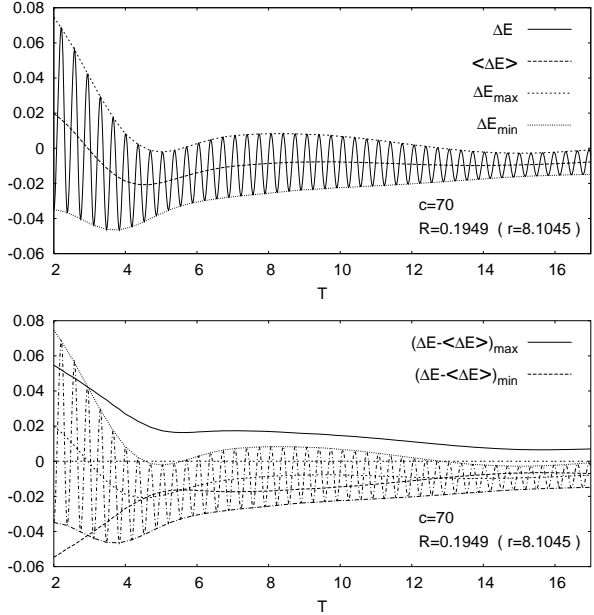}
 } 
\caption{ \label{1eintdR50}   { On the upper plot the difference
$\Delta E = E(T,R)-E_s(R)$, along with the maximum, minimum contours
and the mean values, is plotted for the time interval $2\leq T\leq
17$. On the lower plot the maximum, minimum contours associated with
the separated high frequency oscillations, i.e., with $\Delta E -
\langle \Delta E \rangle$, are shown, where $\langle \Delta E \rangle$
denotes the mean value of the difference $\Delta E = E(T,R)-E_s(R)$.
}}
\end{minipage}
}
\end{figure}

One of the possible methods in monitoring the main features of the
energy transfers is to investigate in more detail the space and time
dependence of the extra energy content of the dynamical monopole.
Fig.\,\ref{1eintdR50} pictures the time dependence of the difference
$\Delta E = E(T,R)-E_s(R)$ along the constant $R$-slice, with
$R=0.1949$, corresponding to $r=8.1045$.
In particular, on the upper part of the figure the time
dependence of $\Delta E$, along with its maximum and minimum contours, together
with the associated average or mean value, are plotted for the time
interval $2\leq T\leq 17$. Notice that, even in case of the considered
intermediate excitation, at the indicated relatively long time period
the mean value of the difference $E(T,R)-E_s(R)$ is still
negative. Looking at this plot it is also clear that the time
dependence of the amplitude of the high frequency oscillations and
that of the mean value should be investigated separately.
For this purpose we investigated separately the time dependence of the mean
value $\langle\Delta E\rangle$, and the time dependence of the amplitude of
the high frequency oscillations $(\Delta E - \langle \Delta E \rangle)_{max}$.
On the lower part of Fig.\,\ref{1eintdR50} the behavior of these
separated variables are shown for the same time period as on the upper
graph.

\begin{figure}[ht]
\unitlength1cm
\centerline{
\begin{minipage}[t]{8.cm}
 \centerline{
  \epsfxsize=8.5cm 
  \epsfbox{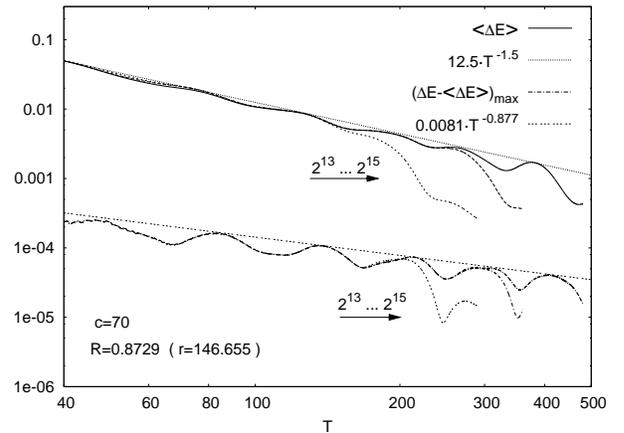}
 } 
\caption{ \label{1eintdR210}   {  The convergence in the {\it log-log}
plots of $\Delta E$ and $(\Delta E - \langle \Delta E\rangle)_{max}$
is demonstrated along the $R$-slice with $R=0.8729$ and for the time
interval $40\leq T\leq 500$ shown. This {\it log-log} plot clearly
manifests also the power law decay of the amplitude of the high
frequency oscillations, $(\Delta E - \langle \Delta E \rangle)_{max}$,
and that of the mean value, $\langle \Delta E \rangle$, with the
particular values $\alpha=0.877$ and $\mu=1.5$.   }}
\end{minipage}
}
\end{figure}

By making use of {\it log-log} plots, see e.g.
Fig.\,\ref{1eintdR210}, for various $R$-slices it  is again
straightforward to recognize a power law time decay that can be
associated with both the amplitude of the high frequency oscillations,
$(\Delta E - \langle \Delta E \rangle)_{max}$, and that of the mean
value, $\langle \Delta E \rangle$. In particular, denoting by
$-\alpha$ and $-\mu$ the two power law exponents  we have that
$\langle \Delta E \rangle \sim T^{-\mu}$ and $(\Delta E - \langle
\Delta E \rangle)_{max} \sim T^{-\alpha}$, respectively.


Table \ref{eintdRt} collects more information about the $R$-dependence
of the values of the power law decaying  exponents $\alpha$ and $\mu$
in case of excitation with $c=70$. Apparently there is a certain
universality in the dying out of the high frequency oscillations,
where the numerical value of $\alpha$ is very close to
$5/6$.  Nevertheless, contrary to our expectations, it turned
out that the  time dependence of the mean value $\langle \Delta E
\rangle$ reflects some sort of non-universal features, i.e., the value of
$\mu$ varies, so it is, in general, different from $1.5$ which had been
associated with the $R=0.8305$-slices in case of the excitations with
amplitudes $c=70$ and $c=280$  (see Figs.\,\ref{eid} and
\ref{eid200b}).

\begin{table} [h]
	  \centering
\begin{tabular}{|c|c|c|c|}
\hline
$r$ & $R$ & $\mu$ & $\alpha$  \\
\hline
\hline
1.0176 & 0.0254 & 0.9 &  0.83333 \\
\hline
2.7244  & 0.0678 &  0.94 &  0.83333   \\
\hline
4.4609 & 0.110169 & 0.9 & 0.828 \\
\hline
6.2471 & 0.1525 & 0.87 & 0.833 \\
\hline
8.1045 & 0.1949 & 0.86 & 0.83 \\
\hline
13.232 & 0.3010 & 1.2 & 0.8333 \\
\hline
19.497 & 0.4068 & 1.5 &  0.83333 \\
\hline
27.822 & 0.5127 & 1.52 &  0.833 \\
\hline
40.088  & 0.6186 & 1.54 & 0.8333 \\
\hline
61.018  & 0.7246 & 1.55 & 0.833 \\
\hline
107.074  & 0.8305 & 1.53 & 0.84 \\
\hline
146.655 & 0.8729  & 1.5 & 0.877  \\
\hline  
\end{tabular}
\caption{ \label{eintdRt}   { The asymptotic time dependence of
$\langle \Delta E \rangle$ and $(\Delta E - \langle \Delta E
\rangle)_{max}$,  along $R=const$ slices, were found to follow the
power law time decay $\langle \Delta E \rangle \sim T^{-\mu}$ and
$(\Delta E - \langle \Delta E \rangle)_{max} \sim T^{-\alpha}$,
respectively.  This table provides the values of $\mu$ and $\alpha$
for the indicated $R$-slices in case of excitation with $c=70$.  }}
\end{table}

It is important to emphasize that by the mere nature of the applied
approximations there is no numerical method which could represent the
full history of the expanding shells of high frequency oscillations
properly. Therefore it is of distinguished importance to know how
strongly the above reported results, concerning the energy transport,
depend on the applied numerical resolution. Is there at all a
meaningful convergence guaranteed, say, at a suitably large value of
$R$ close to the region occupied by the shells of high frequency
oscillations?  Fig.\,\ref{1eintdR210} is to justify that for the
largest value of $R$ referred to in Table\,\ref{eintdRt}, i.e. at the
$R=0.8729$ slice, such a convergence is guaranteed by our numerical
method.

It is important to interpret the above findings in short physical
terms.  For instance, Table \ref{eintdRt} clearly justifies that there
is an inner region with  certain type of universal time decay for the
mean value of the extra energy content which can be characterized by
the value $\mu \sim 0.86-0.9$. 
Essentially, this region corresponds to the core of the monopole.
Since the value of $\mu$ is the
smallest in this inner region the energy transfer remains at a much
higher level here for the entire evolution. 
It is also interesting
that the value of $\mu$ is smaller a bit apart from the centre than
close to the centre which indicates that there remains some energy
bouncing back and forth during the entire evolution in the central
region where the monopole lives.  Then, there is a wide region at the
middle, characterized by the interval $ 0.3 \leq R \leq 0.87 $, where
the mean value of the extra energy content is decaying more rapidly
than elsewhere. Notice that the increase of the exponent
$\mu$ from about $R=0.3$ till $R=0.72$ with the increase of the value
of $R$ indicates that larger and larger fraction of the extra energy
provided by the exciting pulse is stored by the
expanding shells of high frequency oscillations of the Yang-Mills
field.
In fact, this conclusion is also supported by the slower and
slower time decay as the value of $R$ is further increased above
$R=0.72$.

Let us finally investigate the energy content stored in expanding shells of high
frequency oscillation formed by the massive Yang-Mills field represented by
$w$ at large radiuses.
The energy content outside of $\bar R$ is $E(T,1)-E(T,\bar R)$. Since initially
the energy contained in this region is $E_s(1)-E_s(\bar R)$, the energy stored
in the shells is
\begin{equation}
E_{\rm shells}=E(T,1)-E(T,\bar R)-\left[E_s(1)-E_s(\bar R)\right] \, 
\end{equation}
which can be written as
\begin{eqnarray}
E_{\rm shells}&&=\left[E(T,1)-E(\infty,1)\right]
-\left[E(T,\bar R)-E_s(\bar R)\right]\nonumber\\
&&+\left[E(\infty,1)-E_s(1)\right] \ . \label{eshell}
\, 
\end{eqnarray}
Recall (see subsection\,\ref{et1}) that the mean value of the total
energy $E(T,1)$ associated with $T=const$ slices is
approaching its non-zero limit value $E(\infty,1)$ by following a time
decay of the form $T^{-2/3}$. 
As it can be seen from Table \ref{eintdRt}, the second term on the right hand
side of (\ref{eshell}) always dies out faster than the first one.
This relation justifies that the energy stored by the shells of high frequency
oscillations has to approach asymptotically the positive value
$E(\infty,1)-E_s(1)$ from above with the power law decay $\sim T^{-2/3}$.

\subsubsection{The time dependence of the frequency of
  oscillations}\label{freq}  

\begin{figure}[ht]
\unitlength1cm
\centerline{
\begin{minipage}[t]{8.cm}
 \centerline{
  \epsfxsize=8.5cm 
  \epsfbox{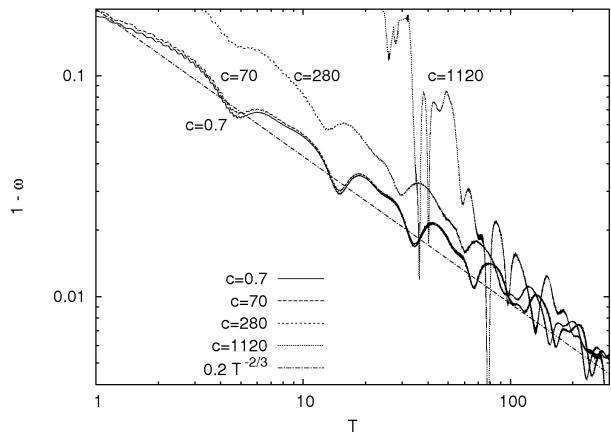}
 } 
\caption{ \label{frcdep}   {  
Time dependence of the oscillation frequency $\omega$ of the field $w$ at
$R=0.02542$, i.e. at $r=1.017$, for four different initial data labelled by the
strength parameter $c$. 
Since the frequencies approach $1$ from below, the value of $1-\omega$ is
plotted logarithmically.
The evolution of the frequency for the weak ($c=0.7$) and intermediate
energy ($c=70$) excitations tends as $T^{-2/3}$ to the limit value $1$.
For the higher energy simulations $1-\omega$ falls off more quickly.
For the $c=1120$ simulation the downwards pointing peaks correspond to moments
where the frequency approaches $1$ closely from below and then gets smaller
again.}}
\end{minipage}
}
\end{figure}

The time dependence of the frequency of the massive $w$ and massless $h$ fields
display markedly different characteristics. Frequency dependence of composite 
quantities such as the energy density will show a mixture of these properties.
As can be seen on Figures
\ref{hd07} and \ref{h280G}, the behavior of the Higgs field $h$ is essentially
the same at each radius $R$, apart from a slowly increasing phase shift going
outwards. But even this time delay is not very significant, since this change
is along outgoing null geodesics. A null geodesic emanating from the center
$R=0$ reach null infinity $R=1$ in a coordinate time interval $\Delta T=1$,
which is small compared to the length of the simulations performed by our
numerical code. The behavior of the $w$ field is qualitatively different, as it
is apparent from Figures \ref{wd07} and \ref{w280}. In the central region the
two fields are strongly coupled and consequently their frequency is the same.
On Fig. \ref{frcdep} the time dependence of the frequency of the $w$ field is
shown for four different initial data at a fixed radius $R=0.02542$,
corresponding to $r=1.017$.
The time dependence of the $h$ field at this radius would yield an essentially
identical figure.

\begin{figure}[ht]
\unitlength1cm
\centerline{
\begin{minipage}[t]{8.cm}
 \centerline{
  \epsfxsize=8.5cm 
  \epsfbox{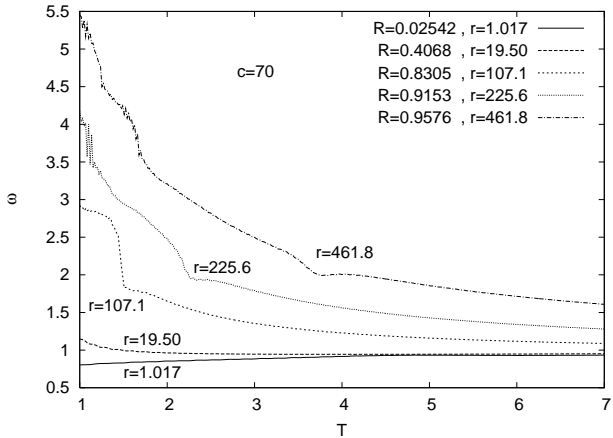}
 } 
\caption{ \label{wrdep}   {
Time dependence of the frequency of the oscillation of the $w$ field at five
different constant radii for a shorter time interval, $1\leq T\leq 7$. The
parameter in the initial data was chosen to be $c=70$. Outside the core of the
monopole the frequency decreases from a peak value reached before $T=1$.
Although the $T<1$ region is not shown because of the high error of our
frequency determination method there, the frequency reaches at least $\omega=10$
for the early stages at larger radii. 
}}
\end{minipage}
}
\end{figure}

\begin{figure}[ht]
\unitlength1cm
\centerline{
\begin{minipage}[t]{8.cm}
 \centerline{
  \epsfxsize=8.5cm 
  \epsfbox{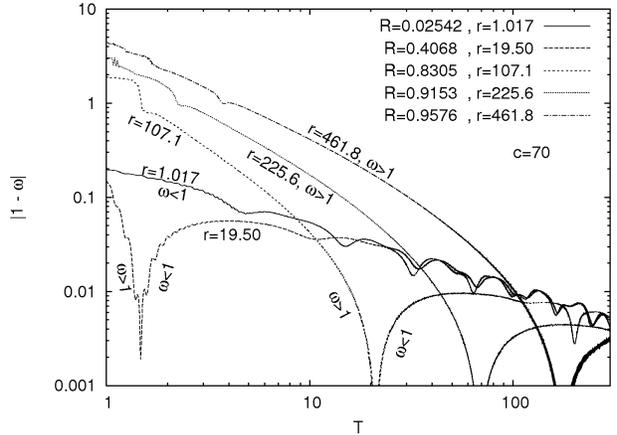}
 } 
\caption{ \label{wrdeplog}   {
Time dependence of the frequency is shown for the same system and same choice 
of radii as on the previous figure, but for a longer time interval, $1\leq T\leq
300$. Since the frequencies tend to $1$, the absolute value of $1-\omega$ is
plotted logarithmically. The downward pointing peaks correspond to moments where
$\omega$ falls below $1$. This happens later for larger radii.
}}
\end{minipage}
}
\end{figure}

The difference between the frequency evolutions of the $w$ and $h$ fields gets
manifested at higher radii, far from the core of the magnetic monopole. At
these distances expanding shell structures appear in $w$, oscillating with high
frequencies. We know in virtue of the results of \cite{fr1} that at a given
radius $R$ the frequency of these shells is the highest just after direct
outgoing geodesics emanating from the initial perturbation region of the
$T=0$ initial hypersurface reach out to the coordinate radius $R$. Of course,
because of the high coordinate velocity of the outgoing null rays, this happens
before $T=1$. It is an interesting result of \cite{fr1} that if $R$ approaches
$1$ then the highest frequency increases without any bound. After the highest
frequency expanding shells have left the radius $R$ the frequency is decreasing
steadily, until it falls to a minimum value below $1$, and then starts to
approach $1$ from below, more or less according to the $T^{-2/3}$ law observed
in the central region. The time evolution of the frequency of the $w$ field for
an initial excitation with $c=70$ is shown at five different radii on Figures
\ref{wrdep} and \ref{wrdeplog}.

\subsubsection{The magnetic charge}

It is known that in a gauge theory, like the one considered in this
paper, by making use of the formal definition applied in Maxwell
theory, a conserved magnetic charge can be defined although the
yielded charge is a Lie-algebra-valued quantity. Hence  a meaningful
definition of magnetic charge requires the use of a gauge
independent specification of what is considered to be the
``electromagnetic content'' of the system. 

Historically two main approaches developed. The first one was
suggested by 't\,Hooft \cite{tH}. It { is based on the use of} the gauge
independent expression \cite{jz}
\begin{equation}
\mathcal{F}_{ab}=\frac{1}{|\psi|}tr(\psi{F}_{ab})
+\frac{i}{g|\psi|^3}tr(\psi[D_a\psi,D_b\psi]),  
\end{equation}
where $|\psi|=tr(\psi\psi)^\frac12$. It can be checked that in terms of the
variable $\mathcal{A}_a=\frac{1}{|\psi|}tr(\psi A_a)$ the above 
``Maxwell'' tensor can be given as 
\begin{equation}
\mathcal{F}_{ab}=\partial_a \mathcal{A}_b-\partial_b \mathcal{A}_a.
\end{equation}
Note, however, that in spite of the elegance of this construction the
main drawback is that according to this definition the magnetic charge
can reside only at the zeros of the Higgs field, which means that in
our case the magnetic charge has to be point-like, it is concentrated
to the origin.  

According to a general expectation the 't\,Hooft-Polyakov magnetic
monopole has the preferable property of being non-singular in contrast
to the Dirac monopole (see e.g. \cite{Goddard}). A  proposal, made by 
Polyakov and Fadeev, fitting 
to this desire, which will also be applied in the rest of this section, is
given by the simple relation \cite{bo}
\begin{equation}
\mathcal{F}_{ab}=\frac{1}{H_0}tr(\psi{F}_{ab}) \ ,
\end{equation}
where $H_0$ is the vacuum expectation value of the Higgs field (see Eq.
(\ref{pot})).

This section is to provide a brief account on the associated
definition of the 
time and location dependent magnetic charge density and the conserved
total  magnetic charge. Note that in case of the applied minimal
dynamical generalization of the static t'Hooft-Polyakov magnetic
monopole no electric charge or electric charge density
are generated. However, in more generic considerations, involving all
the physical degrees of freedom, associated with e.g. the most general
spherically symmetric  excitations of the BPS magnetic monopole, 
non-trivial electric charge density appears, although the vanishing of
the total electric charge is guaranteed  \cite{csfrr}. 

Analogous to the arguments applied in Maxwell theory the magnetic
charge density can be calculated by making use of the magnetic field
strength
\begin{equation}
B_{a}=-\frac{1}{2}\epsilon_{abcd}\,u^{b}\,\mathcal{F}^{cd}
\end{equation}
which is defined with respect to a family of observer represented by
a timelike unit norm 4-velocity field $u^a$. The magnetic charge
density is given then as the 3-divergence of $B_a$ 
\begin{equation}\label{rho}
\rho^{M}={\widetilde{\nabla}}^e B_{e}
\end{equation}
where ${\widetilde{\nabla}}_a$ denotes the covariant derivative operator
associated with the 3-metric induced on the 3-space ``orthogonal'' to the
4-velocity field $u^a$.  

To be more specific and precise, let us choose the timelike unit
4-vector field $u_a$ to be everywhere normal to the $T=const.$
hypersurfaces. Then, for the 
components of $u_a$  
\begin{equation}
u_{\alpha}=\left(\frac{1}{\sqrt{g^{TT}}},\,0,\,0,\,0\right)
\end{equation}
holds, while the induced metric on the $T=const.$ hypersurfaces can be
given as 
\begin{equation}\label{defh}
h_{ab}=g_{ab}- u_{a}\,u_{b}.
\end{equation}
Notice that for any fixed value of $T$ the vector field $u^a$
coincides with the timelike unit norm $n^{(t)}{}^a$ introduced in the
previous section.    

Carrying out all the necessary calculations and combining all the above
relations we get that
\begin{equation}\label{BR}
B_R=-\frac{\Omega}{gR^2}\frac{H}{H_0}(1-w^2),
\end{equation}    
moreover, the expression of the magnetic charge density
$\rho^M$  simplifies to the relation  
\begin{equation}\label{qm}
\rho^M(T,R)=\frac{\Omega^3}{g H_0 R^2}
    {\partial_R\left(H\left[1-w^2\right]\right)},    
\end{equation}
where, in virtue of (\ref{h}) and (\ref{rOm2}), the function $H$
stands for the 
expression $H=h\Omega/R+H_0$. It is important to note that in the
smooth setting the function ${\partial_R\left(H
    [1-w^2]\right)}$ vanishes up to second order in a
small neighborhood of the origin, i.e. there exists a smooth function
$\varphi=\varphi(T,R)$ so that $\partial_R\left(H
    [1-w^2]\right) =\varphi R^2$. This, however,
guarantees that the magnetic charge density $\rho^M$ is regular at the
center as it was anticipated. 

Let us define, again referring to analogies from Maxwell theory,
the magnetic charge associated with a region ${\mathcal{V}}$ in a
spacelike hypersurface $\Sigma$ by the integral  
\begin{equation}
\mathrm{Q}^{M}(\mathcal{V})=\int_{\mathcal{V}}{\tilde{\epsilon}\,\rho^{M}},
\end{equation}
where now $\tilde{\epsilon}$ denotes the 3-volume element of 
the spacelike hypersurface $\Sigma$.

Returning to the previously used specific choice let us choose
${\mathcal{V}}$ to be the ball of radius $R$ centered at the origin on a
$T=const.$ hyperboloidal hypersurface, i.e. ${\mathcal{V}}=
{\mathcal{B}}(T,R)$. Notice, that according to
(\ref{defh}) the components of the 3-volume element 
$\widetilde{\epsilon}$ takes the form given by the relation
(\ref{3vol}). Moreover, in virtue of the Stokes'
theorem and (\ref{rho}) the magnetic
charge associated with the region ${\mathcal{B}}(T,R)$ can be given as
a surface integral on the boundary of ${\mathcal{B}}(T,R)$ 
\begin{equation}
\mathrm{Q}^{M}(T,R)
=\int_{{\mathcal{B}}(T,R)}{\widetilde\nabla^{e}B_{e}\,\tilde{\epsilon}}=
\int_{\partial {\mathcal{B}}(T,R)}{\nu^{e}B_{e}\hat\epsilon}, 
\end{equation}
where $\nu^{a}$ denotes the outward pointing spacelike unit normal
field on the boundary $\partial {\mathcal{B}}(T,R)$ of the region
${\mathcal{B}}(T,R)$, while $\hat\epsilon$ is the volume element associated
with the induced metric $\chi_{ab}$ on $\partial{\mathcal{B}}(T,R)$ which is
given as  
\begin{equation}
\chi_{ab}=h_{ab}+\nu_{a}\,\nu_{b}.
\end{equation}
Since the outward pointing unit normal field $\nu^{a}$ in
the considered special case is proportional to
$\left(\frac{\partial}{\partial R}\right)^a$  we
have for the components of $\nu^{a}$ that 
\begin{equation}
\nu^\alpha=\left(\frac{1}{\sqrt{-h_{RR}}},0,0\right),
\end{equation}
and hence, also that  
\begin{equation}\label{matrixhi}
\chi_{\alpha\beta}=\frac{1}{\Omega^2}\left(
\begin{array}{ccc}
-R^2&0\\
0&-R^2\sin^2{\theta}\\
\end{array}\right),
\end{equation}
which leads to the relation 
\begin{equation}\label{2vol}
\hat \epsilon_{\alpha\beta} = \frac{R^2\sin\theta}{\Omega^2}
\varepsilon_{01\alpha\beta}=\sqrt{|\chi_{_{(2)}}|}(d\theta)_\alpha\wedge
(d\phi)_\beta,
\end{equation}
where $\chi_{_{(2)}}$ is the determinant of the 2-metric
$\chi_{ab}$,
i.e.
\begin{equation}
\sqrt{|\chi_{_{(2)}}|}=\frac{R^2\,\sin{\theta}}{\Omega^2}.
\end{equation}

Combining all the above relations, for the magnetic charge
$\mathrm{Q}^{M}(T,R)$ associated with a ball of radius $R$ and centered
at the origin on a constant $T$ hypersurface we get  
\begin{eqnarray}\label{qqmm}
\mathrm{Q}^{M}(T,R)&=& \int_{-\pi}^{\pi}\int_{0}^{2\pi}
  \nu^{R}B_{R}\sqrt{|\chi_{_{(2)}}|}\,d\theta\,d\phi \nonumber\\ 
&=&  \int_{-\pi}^{\pi}\int_{0}^{2\pi}
  \frac{H(1-w^2)}{gH_0}\sin{\theta}\,d\theta\,d\phi \nonumber\\ 
&=&\frac {4\,\pi }{g}\frac{H}{H_0}\left(1-w^{2}\right).
\end{eqnarray}

The total magnetic charge, $\mathrm{Q}^{M}$, associated with the
entire of a constant $T$ 
hyperboloidal hypersurface, can now be given as 
\begin{equation}
\mathrm{Q}^{M}= \mathrm{Q}^{M}(T,R=1), 
\end{equation}
which, in virtue of (\ref{qqmm}) and since $w$ tends 
to zero exponentially, while $ H/H_0$ tends to one as  
${R\rightarrow 1}$, takes the value  
\begin{equation}
\mathrm{Q}^{M}=\frac {4\,\pi}{g}. 
\end{equation}
Accordingly, the 
total magnetic charge ${Q}^{M}$ is a time independent quantity so that
its value coincides with the value of the topological charge of the
magnetic monopole. Notice also that in virtue of (\ref{qm}) by
exciting the the initially static BPS monopole by 
making use only of a non-zero time derivative of the Yang-Mills variable no
magnetic charge density, and hence, in accordance with the above
conclusion, no magnetic charge is added to the system. 

\begin{figure}[ht]
\unitlength1cm
\centerline{
\begin{minipage}[t]{8.cm}
 \centerline{
  \includegraphics[width=8.5cm,angle=0]{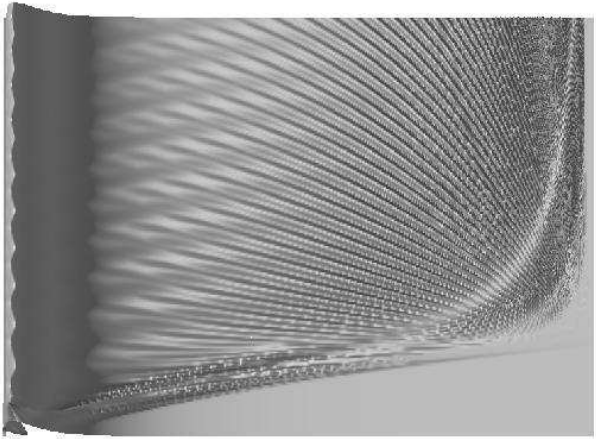}
 }
\caption{\footnotesize \label{rhos}   { The space and time dependence
    of $\tilde \rho^M$, the magnetic charge density associated with
    the shells of radius $R$,  is plotted, for initial data with
    amplitude $c=70$, for the space and time intervals $0\leq R\leq 1$
    and  $0\leq T\leq 4.237288135 $. It is transparent that, even though
    considerably large energy transports occur during the dynamical
    part, the magnetic charge remains concentrated to the central
    region throughout the entire evolution.  Moreover, an intensive
    inward pointing current can also be observed in the distant region
    which quickly returns the lost magnetic charge to the
    center. Accordingly, the amplitude of the
    oscillations is decreasing very rapidly  there.}}
\end{minipage}
}
\end{figure}

On Fig.\,\ref{rhos} the spacetime plot of $\tilde \rho^M$, the
magnetic charge density associated with the shells of radius $R$,
relevant for the  intermediate initial pulse with amplitude $c=70$, is
shown. This quantity, $\tilde \rho^M$, is defined analogously to those
used previously. Accordingly, on any $T=const$ hypersurface the
integral $\int_0^R \tilde \rho^M dR$ is supposed to give the magnetic
charge contained by the ball of radius $R$, and as it can be checked
easily, $\tilde \rho^M= \int_{-\pi}^{\pi}\int_{0}^{2\pi} \rho^M
\sqrt{|h^{(t)}|}\,d\theta\,d\phi$ or, even more directly, $\tilde
\rho^M=4\pi \Omega^3/R^2 \cdot \rho^M =4\pi/(gH_0) \cdot {\partial_R
\left(H \left[1-w^2\right]\right)}$.  In accordance with the facts
that $w$ and $H$ both tend to their asymptotic values, i.e. to zero
and one, respectively, very rapidly, even in case of dynamical
situations a significant part of the total magnetic charge remains in
a small size central region.

\begin{figure}[ht]
\unitlength1cm
\centerline{
\begin{minipage}[t]{8.cm}
 \centerline{
  \epsfxsize=8.5cm 
  \epsfbox{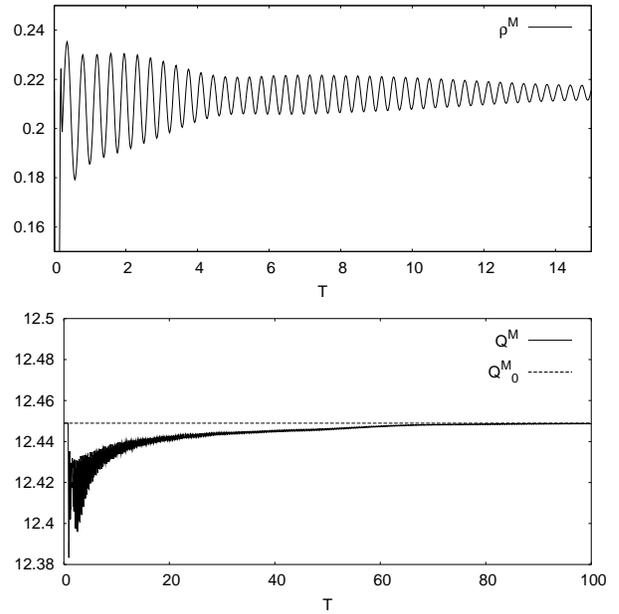}
 }
\caption{ \label{rhoR10}   { On the upper graph the time dependence of
the magnetic charge density $\rho^M$ is shown at the constant $R$-line
with $R=0.067797$ ($r=2.7244$). On lower graph the
time dependence of the magnetic charge contained by the ball of radius
$R=0.83051$ ($r=107.07$), centered at the origin, is shown. Here
${Q^M}_0$ denotes the value of the magnetic charge contained by the
same ball before the arrival of the exciting pulse. Notice that
${Q^M}_0$ is, in fact, the value of the  magnetic charge contained the
same ball in case of the static monopole.}}
\end{minipage}
}
\end{figure}

This conclusion is also supported by the lower graph of
Fig.\,\ref{rhoR10} where the time dependence of $Q^M(T,R)$, the
magnetic charge contained by the ball of radius $R=0.83051$
($r=107.07$) centered at the origin, is shown. Initially the value is
constant until the arrival of the exciting pulse which sweeps out a
small fragment  of the magnetic charge. Nevertheless, the missing
charge returns very rapidly according to the two phase power law
scheme indicated by the graph at the bottom.

\section*{Acknowledgments} 
This research was supported in parts by the OTKA grants T034337,
TS044665, K67942 and the NATO grant PST.CLG.978726. Both of the
authors had been 
Bolyai J\'anos Research Fellows of the Hungarian Academy of Sciences
while I.R. was also a Research Fellow of the Japan Society for the
Promotion of Science.

\section*{Appendix A}

\renewcommand{\theequation}{A.\arabic{equation}}
\setcounter{equation}{0}

The regularity properties (\ref{reorig}) - 
(\ref{reorig2}) of the field variables  
at the origin can be justified by substituting the expansions 
\begin{eqnarray}
&&\hspace{-1.3cm}w(t,r)=\sum_{k=0}^n \frac{1}{k!}
w_k (t)\, r^k +\mathcal{O}_w(r^{n+1}),\label{reg1}\\    
&&\hspace{-1.3cm}H(t,r)=\sum_{k=0}^n \frac{1}{k!}H_k (t)\,
r^k +\mathcal{O}_H(r^{n+1}),  \label{reg2}
\end{eqnarray}
into the field equations and requiring that the coefficients of the
various powers of $r$ vanish identically. The yielded restrictions are
\begin{eqnarray}
w_0(t)&=&1\\
{\partial^k_r w}\vert_{_{r=0}}=w_k(t)&=& 0, 
\end{eqnarray}
for $k$ being odd, and
\begin{eqnarray}
H_0(t)&=&0\\
{\partial^k_r H}\vert_{_{r=0}}= H_k(t)&=& 0, 
\end{eqnarray}
for $k$ being even. It also follows from these relations, along with
(\ref{reg1}) and (\ref{reg2}), that at the origin the time derivatives
${\partial_t w}$ and ${\partial_t H}$ have to vanish throughout the
evolution.  

Note that part of these restrictions can also be deduced in a different
way. In particular, by substituting (\ref{reg1}) and (\ref{reg2}) into the
tetrad components of the energy momentum tensor we immediately get 
that these components cannot be finite at the origin unless the relations
\begin{eqnarray}
w_0(t)&=&1\\
w_1(t)&=&0\\ 
H_0(t)&=&0
\end{eqnarray}
are satisfied.

To get the regularity result at future null infinity we need to
transform the field equations (\ref{ymhe22}) and (\ref{ymhe11}) into a
form suitable to study the asymptotic behavior of the fields at
$\scri^+$. Therefore, instead of the standard coordinates
$(t,r,\theta,\phi)$ we apply coordinates $(u,x,\theta,\phi)$ based on
outgoing null geodesic congruences determined by the relations 
\begin{equation}\label{ctr}
u=t-r \ \ \ {\rm and}\ \ \ x=\frac1r.
\end{equation}
In this frame future null infinity is represented by the $x=0$
hypersurface, moreover, the field equations (\ref{ymhe22}) and
(\ref{ymhe11}) read as 
\begin{eqnarray}
& &\hskip-.4cm x^{2}{\partial ^{2}_x}{w}+2x{\partial_x}{w}+
  2{\partial_x}{\partial_u}{w}=\nonumber\\ &&\hskip+2.2cm
  w\left[\left({w}^{2}-1\right)+\frac{g^2{H}^2}{x^2}
    \right]
\label{ymhe22a} \\
& &\hskip-.4cm x^{3}{\partial ^{2}_x}{H}-2{\partial_u}{H}+
2x{\partial_u }{\partial_x}H= \nonumber\\ &&\hskip+1.3cm
xH\left[2{w}^{2}+\frac{\lambda}{2x^2}(H^2-H_0^2) 
\right]\hskip-.04cm. \label{ymhe11a} 
\end{eqnarray}
Consider now the expansions 
\begin{eqnarray}
&&\hspace{-.9cm} w=\sum_{k=0}^n \frac{1}{k!}
w_k (u)\, x^k +\mathcal{O}_w(x^{n+1}),\label{reg12}\\
& &\hskip-.9cm H=\sum_{k=0}^n \frac{1}{k!}H_k (u)\, 
x^k +\mathcal{O}_H(x^{n+1}),  \label{reg22}
\end{eqnarray}
which are valid in a neighborhood of $x=0$ whenever $w$ and $H$ are at
least $C^n$ functions through $\scrip$, where the notations
$w_k(u)=\partial_x^k w(u,x)\vert_{x=0}$ and $H_k (u)=\partial_x^k
H(u,x)\vert_{x=0}$ have been used. By substituting these expansions
into the field equations (\ref{ymhe22a}) and (\ref{ymhe11a}) we get
the following: Whenever $\lambda\not=0$ 
\begin{eqnarray}\label{resinf}
w(u)&=&0\\ 
H(u)&=&H_0\\
\partial^k_xw&=&w_k (u)=0 \\
\partial^k_xH&=&H_k (u)=0\label{resinf2}
\end{eqnarray}
for all $ 0<k<n$, i.e., in accordance with the results of Winicour
\cite{wini}, both of the fields decay faster than $x^n$ to
their limit  values at $\scrip$. On the other hand, whenever
$\lambda=0$ and $H\hspace{-.08cm}{}_{_\infty}$ is non-zero only the
vanishing of the $x$-derivatives of $w$ (up to 
the order of $n$) is guaranteed  while we get $\partial_u H=0$,
i.e. $H\hspace{-.08cm}{}_{_\infty}$ is independent of $u$, there is no
restriction for $\partial_u\partial_x H$ and, finally,
$\partial_u(\partial^{k}_x H)=0$ for $2\leq k <n$ 
provided that singular behavior of the field at future (or past)
timelike infinite is excluded. Note that the missing of a restriction
on $\partial_u\partial_x H$ is 
in accordance with the following physical picture. The term
$\partial_u\partial_x H$ appears in the energy current expression
thereby whenever $\lambda=0$ the massless Higgs field do transport
energy to $\scrip$. The extent of this energy transport is restricted
only by the initial data and the field equation. 
 
Interesting subcases occur (which would deserve further
investigations) whenever  
\begin{itemize}
\item $\lambda=0$ and $H\hspace{-.08cm}{}_{_\infty}=0$
\item $\lambda\not=0$ and $H_0\not=0$ but
$H\hspace{-.08cm}{}_{_\infty}=0$ 
\item $\lambda\not=0$ and $H_0=0$.
\end{itemize}
Then either or both of the fields are massless and there is no
restriction on the behavior of the actual massless field or fields
at $\scrip$.

The above considerations shows that a YMH system satisfying reasonable
regularity assumptions at $\scrip$ cannot radiate energy to future
null infinity unless either $\lambda=0$, $H_0=0$ or
$H\hspace{-.08cm}{}_{_\infty}=0$. 

\section*{Appendix B}

\renewcommand{\theequation}{B.\arabic{equation}}
\setcounter{equation}{0}

This appendix is to list  expressions we applied to represent
various numerical $R$-derivatives in miscellaneous stencils. The
expressions below always refer to an arbitrary function $f$ on the
$l^{th}$ time slice.  

The first order $R$-derivative relevant for a symmetric
fourth order stencil reads as  
\begin{equation}
\bigl({\partial_R}f\bigr)^l_i \rightarrow
\frac{1}{12\Delta R}\bigl(-f^{l}_{i+2} + 
8f^{l}_{i+1} - 8f^{l}_{i-1} + f^{l}_{i-2}\bigr)\ . 
\end{equation}

The sixth order $R$-derivative relevant for a symmetric
sixth order stencil was approximated as  
\begin{eqnarray}
\hspace{-.7cm}\bigl({\partial^6_R}f\bigr)^l_i & \rightarrow &
\frac{1}{(\Delta R)^6}\bigl( f^{l}_{i+3}  
- 6 f^{l}_{i+2}
\label{6th} 
\\ & &
\hspace{-.5cm}+15 f^{l}_{i+1}
-20f^{l}_{i}  +15 f^{l}_{i-1} -
6f^{l}_{i-2}+ 
f^{l}_{i-3} \bigr).\nonumber  
\end{eqnarray}

We also used a numerical adaptation what would be called as `one sided
derivatives' in analytic investigations. The relevant first order
`left sided derivatives' are approximated in our fourth order stencil as   
\begin{eqnarray}
\hspace{-.4cm}\bigl({\partial_R}f\bigr)^{l}_{_{I_{max}}} & \rightarrow &
\frac {1}{12\Delta R} \bigl(
3f^{l}_{{I_{max}} - 4} \bigr. 
 \\ &&\hspace{-2.1cm}
 \bigl.- 16{f^{l}_{{I_{max}} - 3}}+ 36{f^{l}_{{I_{max}} - 2}}  -
  48{f^{l}_{{I_{max}} - 1}} + 
25{f^{l}_{{I_{max }}}}\bigr)\nonumber
\end{eqnarray}
at $i=I_{max}$ and as  
\begin{eqnarray}
\hspace{-.4cm}\bigl({\partial_R}f\bigr)^{l}_{_{I_{max}-1}} & \rightarrow &
\frac {1}{12\Delta R}\left(
-f^{l}_{{I_{max}} - 4} \right.\\ &
&\hspace{-2.1cm} \left. + 6{f^{l}_{{I_{max}} - 3}} -  
18{f^{l}_{{I_{max}} - 2}} + 10{f^{l}_{{I_{max}} - 1}} +
3{f^{l}_{{I_{max}}}}\right)\nonumber 
\end{eqnarray}
at $i=I_{max}-1$. 

The analogous sixth order `left sided' derivatives
in a sixth order stencil at the spatial gridpoints with indices
$I_{max}$, $I_{max}-1$ and $I_{max}-3$ do not differ from each
other. They are simply yielded by the substitution of $i=I_{max}-3$
into (\ref{6th}).

\vfill\eject

\end{document}